\documentclass{article}
\usepackage[utf8]{inputenc}

\usepackage{subcaption}
\usepackage{amsmath}
\usepackage{url}
\usepackage{multirow}
\usepackage{tabularx}
\usepackage{adjustbox}
\usepackage{verbatim}
\usepackage{microtype}
\usepackage{dirtytalk}
\usepackage{todonotes}
\usepackage{hhline}
\usepackage{amssymb}
\usepackage[section]{placeins}
\usepackage{subcaption}
\usepackage{graphicx}
\usepackage{url}

\usepackage{pdflscape}
\usepackage{afterpage}
\usepackage{capt-of}

\usepackage{bbold}
\usepackage{dsfont}
\usepackage{xcolor}
\usepackage{algorithm}
\usepackage[noend]{algorithmic}

\def\R{\mathbb{R}}

\usepackage{array}
\usepackage{mathtools}

\begin{document}

\title{Knowledge Discovery from Atomic Structures using Feature Importances}

\author{Joakim Linja \and Joonas Hämäläinen \and Antti Pihlajamäki \and Paavo Nieminen \and Sami Malola \and Hannu Häkkinen \and Tommi Kärkkäinen}
\date{\today}

\maketitle

\begin{abstract}

\noindent Molecular-level understanding of the interactions between the constituents of an atomic structure is essential for designing novel materials in various applications. 
This need goes beyond the basic knowledge of the number and types of atoms, their chemical composition, and the character of the chemical interactions. The bigger picture takes place on the quantum level which can be addressed by using the Density-functional theory (DFT).
Use of DFT, however, is a computationally taxing process,
and its results do not readily provide easily interpretable insight into the atomic interactions which would be useful information in material design.
An alternative way to address atomic interactions is to use an interpretable machine learning approach, where a predictive DFT surrogate is constructed and analyzed. The purpose of this paper is to propose such a procedure using a modification of the recently published interpretable distance-based regression method. Our tests with a representative benchmark set of molecules and a complex hybrid nanoparticle confirm the viability and usefulness of the proposed approach. 

\end{abstract}

%==========================================================================
%==========================================================================
\section{Introduction}\label{sec:intro}

Machine learning (ML) has a potential to provide novel insights into quantum-mechanical properties of atomic systems \cite{bib:schutt2017}. 
Better availability of ML methods influences natural sciences in general, by improving material design and characterization, and through accelerating \textit{ab initio} simulations \cite{morgan2020opportunities,raschka2020machine}. 
As stated in \cite{ceriotti2021introduction}, \say{the synergistic application of machine learning and traditional atomistic modeling continues to serve as an accelerator of discovery}. 
Density-functional theory (DFT) is one of the most commonly used methods to simulate the electronic structure of matter. Even if it requires significant amounts of computational power, it is still more efficient than, for example, Hartree-Fock or Quantum Monte Carlo methods. Hence, it is able the generate enough data to enable the use of ML to better understand the properties and dependencies of molecular systems, and to augment and integrate both simulation-based and experimental studies \cite{keith2021combining}. 
Especially in DFT-based computational materials science with applications like materials discovery, drug design, renewable energy, and catalysis, the emergence of the data-driven science paradigm has provided great improvements~\cite{schleder2019dft}.

DFT provides an accurate physics-based calculator able to solve the electronic structure of a molecule or a nanoparticle. 
DFT is, however, computationally taxing, which is why there have been numerous
studies which attempt to form a surrogate of the DFT calculator using machine learning models \cite{bib:whitfield2014,bib:jager2018,bib:zeni2018,bib:panapitiya2018,bib:tkatchenko2020:chemdiscovery,bib:pihlajamaki2020,bib:zhen2021,bib:damore2021,bib:ghosh2022}. 
Construction of a data-driven surrogate model is done through the use of a descriptor. 
Depending on the descriptor, the number of features it provides ranges from tens to thousands \cite{morgan2020opportunities,raschka2020machine,ellingson2020machine,wang2020machine}. 
In addition to being mere surrogates to taxing computations, such datasets of descriptive features generated from atomic configurations can provide a fertile ground to explore the possibilities of automated methods that weigh and compare the necessity and relative importance of each of the descriptive features with respect to the physical behavior of materials. 

Knowledge discovery, in general, is a stepwise process comprising data generation and selection, model or pattern construction, and interpretation of the constructed model to provide knowledge useful in an application area \cite{fayyad1996kdd,rotondo2020evolution}. 
In this paper, we examine data of atomic structures generated using DFT and descriptors. 
We then construct a regression model that predicts DFT energies from the descriptors. Finally, useful knowledge regarding atomic interactions comes from examining feature importance scores that are obtained using our proposed algorithm.

Interpretable machine learning is a field of research, where improved understanding of the working logic of machine learning models and algorithms is addressed and advanced \cite{adadi2018peeking}. 
This can be done in a model-agnostic (MA) or model-specific (MS) way, where the former refers to the use of separate explanation techniques and the latter means direct interpretation of the actual model \cite{burkart2021survey}. 
An example of a model-agnostic approach to better understand gold nanocluster synthesis was given in \cite{li2019deep}, where the decision tree technique was used to extract rule-based knowledge from a graph convolutional neural network model. 
Indeed, the MA approach might be the only possible one if we are using completely black box models and algorithms provided as components or services, whereas the MS approach is viable if we can access and analyze the ML model itself. 
If applicable, the MS interpretation can be more transparent and accurate, thus providing more potential for novel insight and new discoveries \cite{roscher2020explainable}.

A popular technique for interpreting an ML surrogate is the post-hoc investigation of the saliency of features, i.e., evaluation of feature importances \cite{burkart2021survey, WojtasNIPS2020,arrieta2020explainable}.
This technique is behind the most well-known MA techniques like LIME and SHAP \cite{saarela2021comparison,saarela2021explainable}. 
For the distance-based learning machines proposed in \cite{bib:desouza2015,bib:karkkainen2019:EMLM,bib:hamalainen2020}, the model-specific feature importance formulation Mean-Absolute-Sensitivity (MAS) was proposed and thoroughly experimented in \cite{bib:linja2023}.
Indeed, these distance-based methods can be used to construct deterministic and accurate regression models for general and atomic datasets, which scale to high-dimensional feature and target spaces, are tolerant against noisy and irrelevant features, have theoretical guarantees through the universal approximation results, and can provide better generalization results compared to off-the-shelf deep neural networks \cite{%
bib:pihlajamaki2020,%
bib:karkkainen2019:EMLM,%
bib:hamalainen2020,%
bib:linja2023,%
bib:linja2020%
}.
Furthermore, as noted in \cite{bib:linja2023}, calculation of feature scores to represent their importances can be seen as two sides of the same coin: one to select features and the other to perform knowledge discovery.

In this paper, we examine two different kinds of atomic structures: molecules and a protected metal nanocluster.
We study the output of the Many-body Tensor Representation (MBTR) descriptor through the use of a distance-based feature selection algorithm, showing that features can be removed while simultaneously providing information on the studied atomic structures as described by the descriptor. 

The purpose of this article is three-fold:
\begin{description}
\item{$i)$} we extend our earlier feature selection method by proposing a new rule to determine the most important feature scores;
\item{$ii)$} we extend the comparative assessment of the MAS-based method from \cite{bib:linja2023} by considering new datasets;
\item{$iii)$} we provide a proof-of-concept on how the developed feature scoring and ranking method can be used to extract novel knowledge from atomic structures like molecules and hybrid nanoparticles.
\end{description}
To reach these goals, we employ a representative set of atomic structures arising from benchmark molecular datasets and one particular hybrid nanoparticle dataset, which serve both as validation data of the proposed approach and as the proof-of-concept in the atomic structure realm.

The main contributions of this paper are the proposed feature selection algorithm and the showcase of applying it to data from atomic structures and analysis of the results. 
The analysis reveals that the algorithm is able to detect the most meaningful features, which agree with chemical intuition, from the vast amount of information encoded into MBTR. These features are directly related to the bonding nature of the systems, thus they could be used to distinguish specific characteristics of the molecules, nanoparticles or other atomic systems. 

In Section~\ref{sec:Methods} we describe the theoretical background and present the proposed feature selection algorithm. 
Section~\ref{sec:experimentalsetting} presents the datasets and experimental setup. 
This is followed by the results in Section~\ref{sec:results}. 
Finally, the paper is concluded in Section~\ref{sec:conclusions}. 

%==========================================================================
%==========================================================================
\section{Methods}\label{sec:Methods}

In this section, we provide the relevant information on the existing and new methods used in the experiments. We consider the regression problem, where it is assumed that, for a set of $N$ atomic structures $\{ A_i \}_{i=1}^N$, their potential energies have been calculated using DFT. 
This desired regression output data for a surrogate is denoted with $\{ y_i \}_{i=1}^N$. 

%==========================================================================
\subsection{Descriptors and Many-body Tensor Representation}\label{ssec:MBTR}

Descriptors are a way to translate an atomic structure into a format understood by a machine learning model \cite{bib:musil2021}. 
For a descriptor to be functional, it needs to be invariant to the translation and rotation of the described system as well as to the permutations of atom listing \cite{bib:himanen2020}. 
It also needs to be unique for a given structure and continuous, as the descriptor should be able to capture even the smallest changes of the compounds \cite{bib:himanen2020}. 

Descriptors are divided into two main categories, local descriptors and global descriptors \cite{bib:musil2021}. 
Local descriptors focus on describing the environment of a single atom, whereas global descriptors simultaneously describe the entire atomic structure. 
Another way to categorize descriptors is whether or not they are able to handle periodic environments \cite{bib:himanen2020}. 

Musil et al. \cite{bib:musil2021} presented the families of descriptors as atom density fields, potential fields, symmetrized local fields, atom centered distributions, sharp and smooth density correlation features, internal coordinates, atomic symmetry functions, permutation invariant polynomials, distance histograms, sorted distances, sorted eigenvalues and molecular graphs. 
The focus in this paper is on a global, non-periodic (in this case), distance histogram-family descriptor. 

Many Body Tensor Representation (MBTR) by Huo et al. \cite{bib:huo2022} is a global descriptor, which relies on Gaussian broadened geometric properties (distances and angles) grouped according to the atomic numbers. 
It is invariant with respect to translations, rotations, and permutations, which are crucial properties of a descriptor. Without these properties, for example, the indexing of atoms in an atomic system has an effect on the training of the model. 

MBTR has description settings known as ''k1'', ''k2'', and ''k3'' which refer to the level of details taken into account in the
atomic interactions. 
The representation is based on the following density distributions as presented by Himanen et al in \cite{bib:himanen2020} (with $Z$ denoting an atomic number):

\begin{itemize}
    \item k1: the number of elements,
    \begin{align*}
        \text{F}_1^{Z_1} (x) &= \sum_l^{|Z_1|} {w_1^l} D_1^l(x), \ \text{where}\\
        D_1^l(x) &= \frac{1}{\sigma_1 \sqrt{2 \pi}}e^{-\frac{(x -g_1(Z_l))^2}{2 \sigma_1^2}}, \\
        g_1(Z_l) &= Z_l,\ \text{the atomic number,} \\
    \end{align*}
    \item k2: inverse distances between each element pair,
    \begin{align*}
        \text{F}_2^{Z_1,Z_2} (x) &= \sum_l^{|Z_1|} \sum_m^{|Z_2|} {w_2^{l,m}} D_2^{l,m}(x), \ \text{where}\\
        D_2^{l,m}(x) &= \frac{1}{\sigma_2 \sqrt{2 \pi}}e^{-\frac{(x -g_2(\mathbf{R}_l,\mathbf{R}_m))^2}{2 \sigma_2^2}}, \\
        g_2(\mathbf{R}_l,\mathbf{R}_m) &= \frac{1}{\left| \mathbf{R}_l-\mathbf{R}_m \right|} \text{, and}\\
    \end{align*}
    \item k3: angle between each element triple,
    \begin{align*}
        \text{F}_3^{Z_1,Z_2,Z_3} (x) &= \sum_l^{|Z_1|} \sum_m^{|Z_2|} \sum_n^{|Z_3|} {w_3^{l,m,n}} D_3^{l,m,n}(x), \ \text{where}\\
        D_3^{l,m,n}(x) &= \frac{1}{\sigma_3 \sqrt{2 \pi}}e^{-\frac{(x -g_3(\mathbf{R}_l,\mathbf{R}_m,\mathbf{R}_n))^2}{2 \sigma_3^2}}, \\
        g_3(\mathbf{R}_l,\mathbf{R}_m,\mathbf{R}_n) &= \cos \angle \left( \mathbf{R}_l-\mathbf{R}_m, \mathbf{R}_n-\mathbf{R}_m \right) . \\
    \end{align*}
\end{itemize}
Here $\mathbf{R}_i$ denotes the position of an atom $i$. 
The weighting functions $w_k$ can be used to prioritize features if desired \cite{bib:himanen2020}. 
In the context of this paper, $w_1=1$ and, for $w_2$ and $w_3$, we used the exponential weighting functions which are the default settings given by the DScribe implementation of MBTR \cite{bib:himanen2020}. 
The described three feature sets can be expressed together as $\mathbf{F}_{1,2,3} = \langle \text{F}_1^{Z_1}, \text{F}_2^{Z_1,Z_2}, \text{F}_3^{Z_1,Z_2,Z_3} \rangle$. 
The column vector $\mathbf{F}_{1,2,3}$ forms the descriptor used in this paper. 
Put together, $\mathbf{F}_{1,2,3}$ forms high-dimensional data. 
Ranging from $1100$ features to $5400$ features with the used datasets. 

\begin{figure}[b!]
    \centering
    \includegraphics[width=\textwidth]{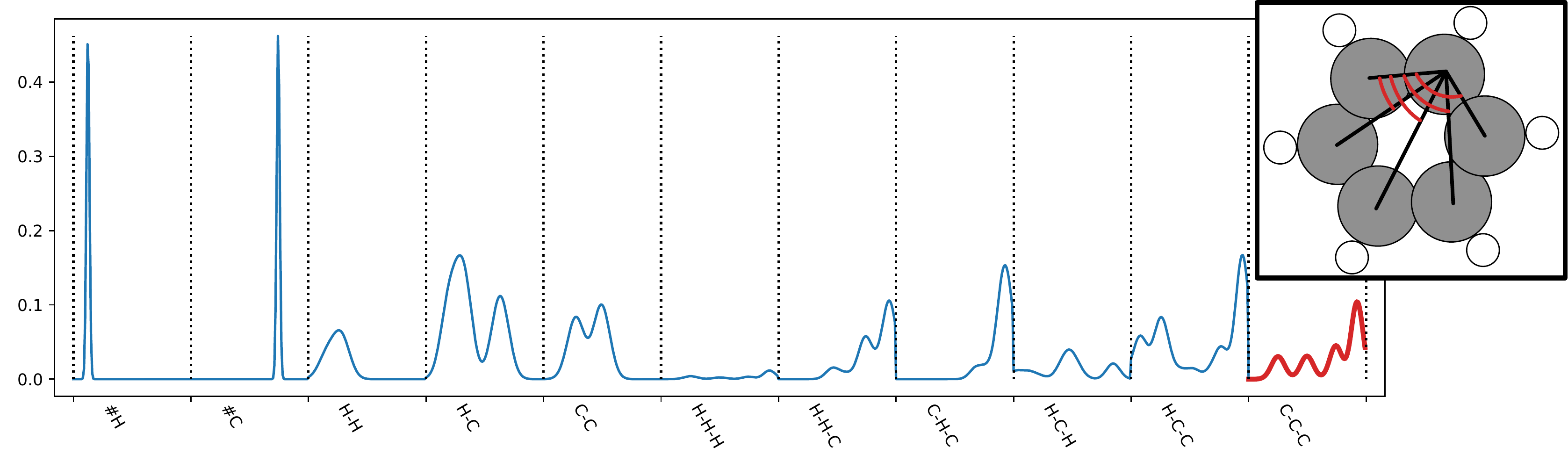}
    \caption{Example of an MBTR descriptor (k1+k2+k3) for Benzene. Each individual atomic interaction is labeled and given dashed vertical lines as borders. The first two, \#H and \#C, represent k1, the following three (H-H, H-C, and C-C) represent k2, and the remaining interactions represent k3.}
    \label{fig:MBTR_Example}
\end{figure}

Used together, $\mathbf{F}_{1,2,3}$ describes the number of atoms in an atomic structure, their distances to each other and the angles between them, forming an extensive description of the structure. 
This description can then be used to study both feature selection and feature importance-based knowledge discovery. 
In Figure~\ref{fig:MBTR_Example}, we provide an example of using MBTR with the dataset \emph{Benzene} (presented in Section~\ref{ssec:datasets}), including also a visualization of the molecule. 
The x-axis in the figure groups together the different interaction types as presented by the MBTR and the y-axis depicts
\emph{mean MBTR}: mean over the observations of the whole \emph{Benzene} dataset. 
The interaction C-C-C has been highlighted in red in Figure~\ref{fig:MBTR_Example}. 
This interaction is taken as an example of how the descriptor is formed. 
All possible angles between carbon atoms have been marked with red lines and angle notations in the inset molecule image of Figure~\ref{fig:MBTR_Example}. 
For each descriptor, these angles are calculated for each carbon atom individually, and the combination then forms the C-C-C line highlighted with red. 
The way to read the notation is that if we were to have an angle A-B-C between atoms A, B and C, it would read as the angle between vectors B-A and B-C.

%==========================================================================
\subsection{Extreme Minimal Learning Machine}\label{ssec:EMLM}

\emph{Extreme Minimal Learning Machine} (EMLM) is a distance-based machine learning model proposed by Kärkkäinen in 2019 \cite{bib:karkkainen2019:EMLM}. 
EMLM is a combination of Extreme Learning Machine (ELM) \cite{bib:huang2004,bib:huang2006} and Minimal Learning Machine (MLM) \cite{bib:desouza2015,bib:hamalainen2020}, combining the distance-based feature map 
of MLM to the way a regularized least-squares problem is solved in ELM. 
EMLM works by first constructing a distance matrix
\begin{equation}\label{eq:DistMat}
(\mathbf{H})_{ij} = || \mathbf{r}_i - \mathbf{x}_j ||_2, i = 1,\ldots, m; j = 1,\ldots, N,
\end{equation}
between observations $\mathbf{x}_j$ and a set of reference points $\mathbf{r}_i$. Here the reference points $\mathbf{R} = \{ \mathbf{r}_i \}_{i=1}^m$ are selected from the given set of $N$ observations $\mathbf{X} = \{ \mathbf{x}_i \}_{i=1}^N$, i.e. $\mathbf{R} \subseteq \mathbf{X}$, and this selection procedure includes the only metaparameter of the method: number of reference points $m$. A full EMLM would use all observations as reference points (providing a parameter-free method) but a typical choice would be to apply a strategy, such as RS-maximin, which enables a smaller number of reference points to cover the $\mathbf{X}$ in a representative manner \cite{bib:hamalainen2018,bib:hamalainen2020}. 

In training of EMLM, the weights $\mathbf{W}$ are calculated using the distance matrix $\mathbf{H}$, similarly to the classical ridge regression formula in linear regression:
\begin{equation}\label{eq:EMLMEqu}
    \mathbf{W} \left( \mathbf{H}\mathbf{H}^T + \frac{\alpha N}{m}\mathbf{I}  \right) = \mathbf{y} \mathbf{H}^T,
\end{equation}
where $\mathbf{y}$ is the vector of DFT-outputs and $\frac{\alpha N}{m}\mathbf{I}$ the least-squares (ridge regression) regularization term. 
In the equation, $\alpha = \sqrt{\epsilon}$ where $\epsilon$ refers to machine epsilon. 
After the weights $\mathbf{W}$ have been obtained from \eqref{eq:EMLMEqu}, the model's prediction of a new input vector $\tilde{\mathbf{x}}$ is calculated by forming the distance matrix between the reference points and $\tilde{\mathbf{x}}$ according to \eqref{eq:DistMat} and then multiplying this with the weights.

The strength of EMLM and other such distance-based models lies in their simplicity of use and robustness. 
Most importantly, a large and versatile pool of experiments confirm their repeated tendency not to overlearn \cite{bib:karkkainen2019:EMLM,bib:hamalainen2020,bib:linja2020,bib:pihlajamaki2020,bib:linja2023}. 
From the DFT-surrogate construction perspective, this makes these techniques very different from the popular deep learning methods, which require tedious and extensive selection, tuning, and testing of architecture and metaparameters to manage model complexity and to prevent overlearning \cite{koutsoukas2017deep,li2021machine}.

%==========================================================================
\subsection{Feature scoring methods}\label{ssec:FScore}

The more complex a measured phenomenon is, the more features are usually required for a surrogate. 
Events in real world are quite often complex, with many variables. 
This produces datasets with large numbers of features. 
However, the assumption is that not all of the measured variables produce features which are actually necessary. 

Kohavi and John presented definitions for strong and weak relevance in \cite{bib:kohavi1998}, and noted that the relevance of a feature may not be directly correlated with the optimality of the feature (which clearly is related to what kind of data-driven model is to be constructed). 
Feature selection algorithms rely on feature scoring in order to assess their relevance or importance. Based on how the scores and/or determination of feature importances is realized, the actual feature selection algorithms can be divided into three main categories: $i)$  Filters execute a rule by which they select a feature subset without data-driven model construction. $ii)$  Wrappers, which use the machine learning model to assess the importance of features. and $iii)$ Hybrids, which are in one way or another a combination of a filter and a wrapper (like decision trees and random forest). 

Based on the extensive experiments presented in \cite{bib:linja2023}, only the best two feature scoring, ranking, and selection algorithms are used and experimented further in this paper: \emph{Mean Absolute Sensitivity} and \emph{Spearman R}.

\subsubsection*{Feature scoring using Mean Absolute Sensitivity}\label{ssec:MAS}

The partial derivative of a neural network’s output with respect to its input to measure the input sensitivity was proposed in \cite{dimopoulos1995use}. This technique was enlarged to feature selection in \cite{karkkainen2015assessment}. In order to generate an image-specific saliency map for visual interpretation of a convolutional neural network classifier, the input-output sensitivity was independently rediscovered and proposed in \cite{simonyan2013deep}.

Formally, as depicted in \cite{KarkPN2022}, the derivative of a data-driven model $M$ with respect to its features $\frac{\partial M}{\partial \mathbf{x}}$ to assess feature importances emerges from the classical Taylor's formula. Explicit formulae for the distance-based EMLM and also for a deep, feedforward neural network to calculate a model's derivatives for an observation were given in \cite{bib:linja2023}.
Such calculation provides a pointwise information so that in order to assess the global importance, one can take mean of the absolute sensitivities (MAS) over the data \cite{karkkainen2015assessment}: 
\begin{equation}\label{eq:mas}
MAS = \frac{1}{N} \sum_{i=1}^{N} \left| \frac{\partial M}{\partial \mathbf{x}_i} \right| . 
\end{equation}
The actual formulae for the EMLM, for $\tilde{\mathbf{x}} = \mathbf{x}_i$ were given in \cite{bib:linja2023}:
\begin{equation}\label{eq:MASdetails}
\frac{\partial M}{\partial \tilde{\mathbf{x}}} = \mathbf{W} \mathbf{D}^T ,
\quad \text{where} \  
\mathbf{d}_i = \frac{\mathbf{r_i} - \tilde{\mathbf{x}}}{\max (\alpha, \| \mathbf{r_i} - \tilde{\mathbf{x}} \|)}, i = 1, \ldots, m .
\end{equation}

MAS can be calculated for any data-driven model which can be differentiated with respect to its features. 
For regression problems, the calculated MAS value of a feature estimates its importance in the construction of model $M$ compared to other features, allowing thus MAS to function as a straightforward score for ranking and knowledge discovery purposes. 

%==========================================================================
\subsubsection*{Correlation-based feature scoring using Spearman R}\label{ssec:theory:FS}

The premise of correlation-based feature assessment is that features can be ranked according to their correlation to the output, especially in regression problems \cite{bib:hall1999}. 
A feature may be positively or negatively correlated, or neutral in respect of the output \cite{bib:bell2000}. 
In general, features can be categorized into having a strong relevance or a weak relevance \cite{bib:kohavi1998}. 

Spearman rank correlation coefficient is a statistical method, which is a non-parametric way to calculate the correlation between two groups of data \cite{bib:spearman1904}. 
In addition, Spearman R does not assume any distributions, which is important since a dataset cannot be expected to have normal distribution. The classical formula, for the correlation coefficient vector $\mathbf{\rho}$, reads as
$$
\mathbf{\rho} = \frac{\sum_{i=1}^N (\mathbf{x}_i - \mathbf{\bar{x}}) (\mathbf{y}_i - \mathbf{\bar{y}})}{\sqrt{\sum_{i=1}^N (\mathbf{x}_i - \mathbf{\bar{x}})^2 \sum_{i=1}^N (\mathbf{y}_i - \mathbf{\bar{y}})^2}} .
$$
Note that if no preprocessing is assumed then, similarly to \eqref{eq:MASdetails}, the denominator of the correlation should be safeguarded against the case of a constant feature.

%==========================================================================
\subsection{The developed algorithm}\label{ssec:FSalg}

\begin{figure}[ht!]
    \centering
    \includegraphics[width=\textwidth]{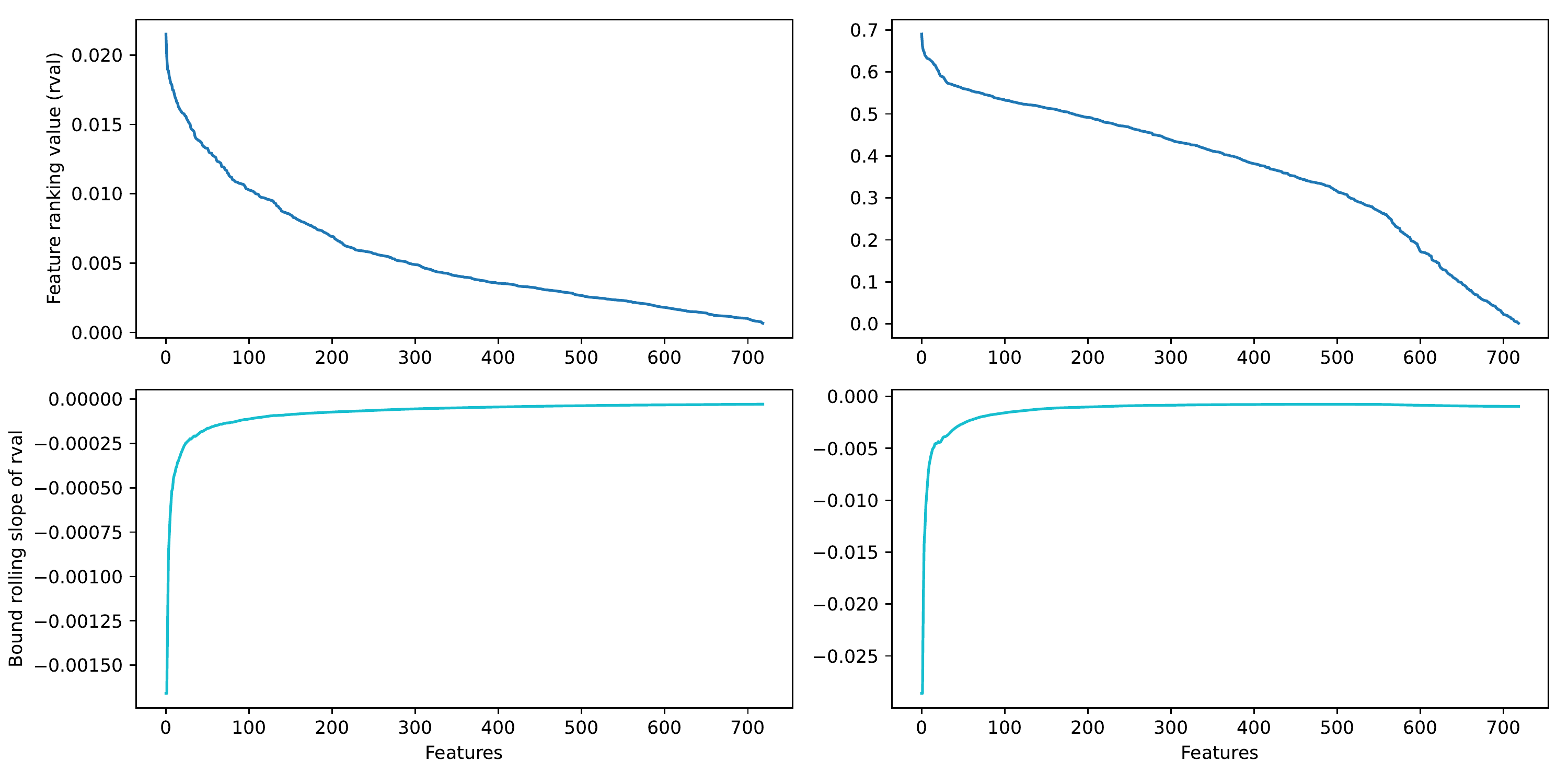}
    \caption{The ranking values of Au$_{38}$ data calculated with MAS and with Spearman R as well as the next step of the proposed feature selection algorithm to point out that the shape of the ranking values produces similar curve in both cases}
    \label{fig:reasonforfsalg}
\end{figure}

The proposed feature scoring, ranking, selection algorithm is an extension on the idea of the one-shot wrapper presented by Linja et al. in \cite{bib:linja2023}. 
The development was started by the discovery that there are cases when the features of a dataset are ranked and sorted, the curve formed by the ranking values is not suitable for a kneepoint detection. 
The problematic situation is illustrated in Figure~\ref{fig:reasonforfsalg}. 

\begin{figure}[ht!]
    \centering
    \includegraphics[width=\textwidth]{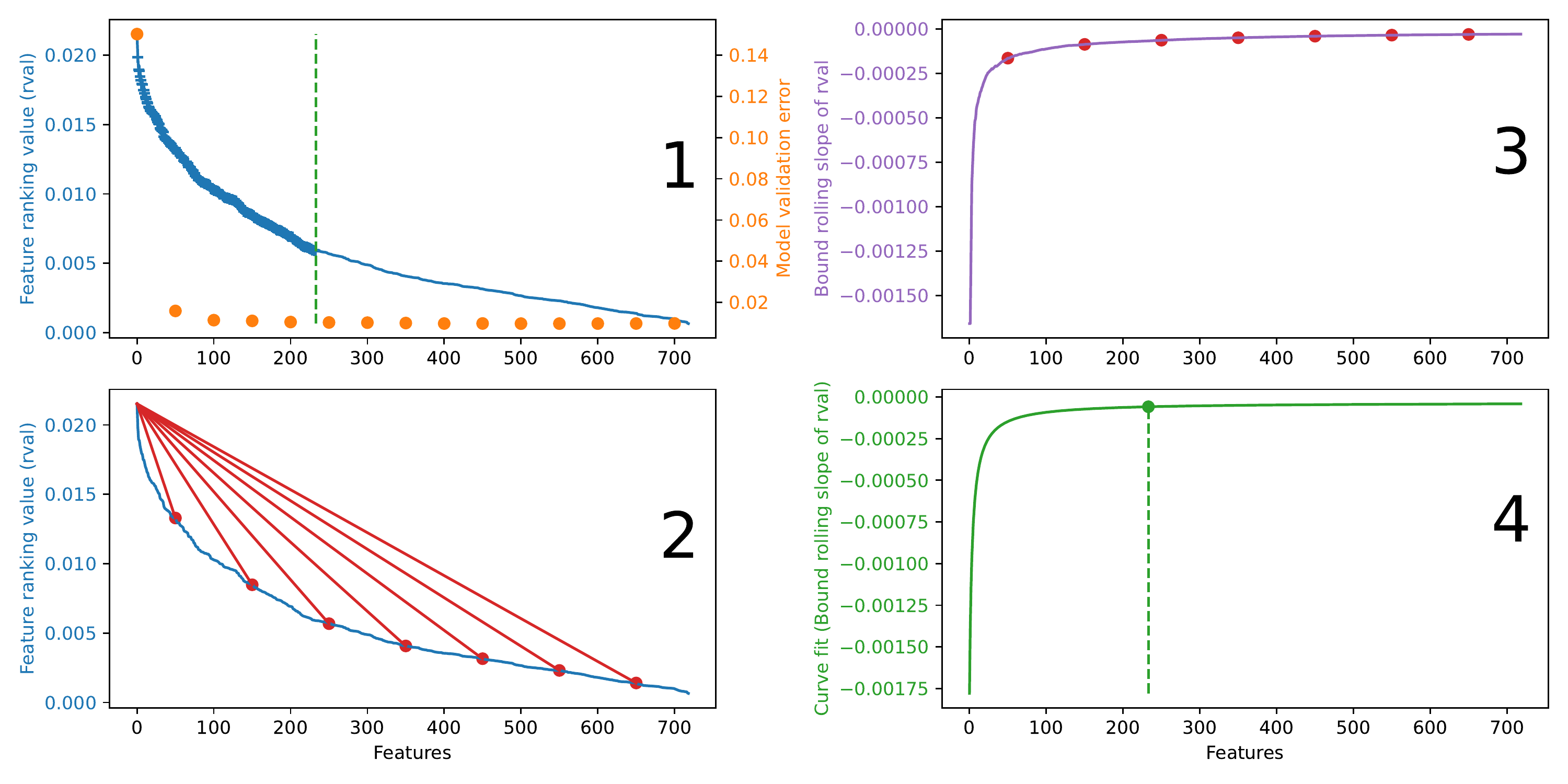}
    \caption{Illustration of the function of the proposed feature selection algorithm}
    \label{fig:fsillustration}
\end{figure}

The proposed algorithm aims to fix the issue by calculating the slope of each line between the highest ranking value and the other ranking values. 
It then performs a curve fit to the slope values and uses the fitted curve to determine the selected features. 
The algorithm is presented in Algorithm~\ref{alg:FID}. 
It's function is illustrated in Figure~\ref{fig:fsillustration}. 

Figure~\ref{fig:fsillustration} uses the Au$_{38}$ dataset as an example, showing the ranking values of features of MBTR k2 as calculated by MAS and set to descending order in subfigure 1. 
It also shows the validation errors of the model as well as the final cutoff point given by the last step in subfigure 4. 
Subfigure 2 shows the bound rolling slope which forms the curve in subfigure 3. 
Finally, subfigure 4 shows a curve fit of the curve in subfigure 3 and the cutoff point. 

\begin{algorithm}[b!]
\caption{Feature Importance Detector - FID}
\label{alg:FID}
\begin{algorithmic}[1]
\REQUIRE Input data $\{\mathbf{x}_j \in \R^n \, | \, j=1,\ldots,N\}$, target data $\{y_j \in \R \, | \, j=1,\ldots,N\}$, cutoff value $\{k \in \R \, | \, 0 < k < 1\}$
\ENSURE 
Feature score vector $\mathbf{v}$, Ranking permutation $\mathbf{p}$, and indices of the most important features $\mathcal{F}$
    \STATE Score the $n$ features of the input data by using a feature scoring algorithm \textemdash here Spearman R or MAS 
    \STATE Sort the feature scoring values $\{v_i\}_{i=1}^n$ into descending order and keep permutation function $p(i)$ which allows to return from sorted indices to original ones
    \STATE Calculate a slope value for each feature: $s_i = \frac{v_i-v_1}{i-1}, i = 2,\ldots,n$
    \STATE Estimate $a,b,c,d \in \R$ by fitting the curve $C(i) = -\frac{a}{bi+c}+d$ to the points $\{s_i\}$ 
    \STATE Define total increase of $C(i)$ denoted as $C^* = C(n)-C(1)$
    \STATE Return feature set $\mathcal{F} = \{p(i)\, |\, C(i) < C(n) -k\, C^*, i = 1,\ldots,n \}$
\end{algorithmic}
\end{algorithm}

%==========================================================================
%==========================================================================
\section{Experimental Setting}\label{sec:experimentalsetting}

%==========================================================================
\subsection{Datasets}\label{ssec:datasets} %\\

In this study, we used the following datasets of the atomic structures: all benchmark data from sGDML (Symmetric Gradient Domain Machine Learning) \cite{bib:chmiela2017,bib:schutt2017,bib:chmiela2018} and the hybrid nanoparticle dataset \emph{Au38QT}, which originates from the simulations by Juarez-Mosqueda et al. \cite{bib:rosalba2019}. 
Among the used datasets, \emph{Au38QT} is of special interest because of its higher complexity and due to the interest nanoscience community and machine learning community are giving to nanoclusters, as well as our own interest in it \cite{bib:pihlajamaki2020,bib:linja2020}. 
Its structure was first found experimentally by Qian et al in 2010 and an isomer for it was found by Tian et al in 2015 \cite{bib:qian2010:au38structure,bib:tian2015:nanoparticles}. 
As a monolayer protected cluster (MPC), it holds high levels of potential applications, such as biolabeling, catalysis, medicine, solar energy and display panels \cite{bib:tsukuda2015protected}. 

The datasets are presented in Table~\ref{tab:datasets}. 
Example images of the molecule or cluster of each dataset is presented in Appendix~A. 
Each dataset is based on either a real--world molecule or a nanoparticle. 
The observations in each have been gained by simulating the atomic object in DFT-based molecular dynamics simulation. 
The target variable in each dataset is the potential energy of an observations configuration. 

\begin{table}
    \centering
        \caption{Datasets used in this study. The number of features is determined by the used descriptor, described in Section~\ref{ssec:MBTR} and Section~\ref{ssec:experimentalSetup}.}
    \begin{tabular}[htbp]{@{}llll@{}}
        \hline
        Dataset & \# Observations & \# Features &  \# Elements \\
        \hline
        Au38QT              & 28761 & 5400 & H72,C24,S24,Au38 \\
        Benzene2018         & 49863 & 1100 & H6,C6 \\
        DocosahexaenoicAcid & 69753 & 2700 & H32,C22,O2 \\
        AlanineTetrapeptide & 85109 & 5400 & H22,C12,N4,O4 \\
        Azobenzene          & 99999 & 2700 & H10,C12,N2 \\
        Paracetamol         & 106490 & 5400 & H9,C8,N1,O2 \\
        Uracil              & 133770 & 5400 & H4,C4,N2,O2 \\
        Aspirin             & 211762 & 2700 & H8,C9,O4 \\
        Salicylic Acid      & 320231 & 2700 & H6,C7,O3 \\
        Naphthalene         & 326250 & 1100 & H8,C10 \\
        Toluene             & 442790 & 1100 & H8,C7 \\
        Ethanol             & 555092 & 2700 & H6,C2,O1 \\
        \hline
    \end{tabular}
    \label{tab:datasets}
\end{table}

%==========================================================================
\subsection{Experimental setup}\label{ssec:experimentalSetup} %\\

In the molecular dynamics simulation data, each atom configuration was converted into a set of descriptors using MBTR (k1+k2+k3). 
Using all three types of MBTR descriptors produces a vector which describes the elements, distances and angles of each atom configuration. 
Each dataset was split into subsets according to their target variable using distribution-optimal folding, 
DOP-SCV \cite{moreno2012study,karkkainen2014cross}. 
The number of subsets was set to be $5$ for datasets with less than $250000$ observations and $10$ for the rest. 
For each subset, constant features and duplicate observations were removed. 
In addition, each subset was minmax-scaled to the interval $[0,1]$. 
Then, the feature importance detection algorithm, Algorithm~\ref{alg:FID}, was performed on each data subset.
The following variables were measured: 
\begin{itemize}
    \item baseline validation RMSE of a data subset,
    \item the same after feature selection has reduced the number of features in the subset and
    \item whether or not a feature was selected and if removed, the point in the algorithm when it was removed. 
\end{itemize}
In order to focus the experiments on model's generalization instead of its scalability, We performed cross-validation in reverse fashion, such that one subset was used as a training set and the rest of the subsets as validation sets. 
The RMSE error was calculated in electron volts (eV). 

We evaluated the feature selection capability of Algorithm~\ref{alg:FID} using two different feature scoring algorithms, \emph{MAS} and \emph{Spearman R}, and three different cutoff values: $10^{-1}$, $10^{-2}$ and $10^{-3}$. 
In addition, we extracted information from MBTR regarding the relevance of each atomic interaction via a \emph{term frequency -- inverse document frequency}-type measurement \cite{leskovec2020mining}, denoted here as ''interaction relevance'' $\iota$: 

\begin{equation}\label{eq:interactionRelevance}
    \iota (\text{F}_{k}^{Z_1,\ldots,Z_k}) = \frac{1}{N_\text{D}} \sum_{d \in D} \frac{N_{\text{FS}}(d)}{N_\text{res}}, 
\end{equation}
where $\text{F}_{k}^{Z_1,\ldots,Z_k}$ is the atomic interaction (as defined by MBTR, for example, H-C) $N_\text{D}$ is the number of datasets in which  $\text{F}_{k}^{Z_1,\ldots,Z_k}$ appears in, $D$ is a set of datasets in which $\text{F}_{k}^{Z_1,\ldots,Z_k}$ appears in, $d \in D$ refers to a dataset, $N_{\text{FS}}$ is the number of features that remains of the described $\text{F}_{k}^{Z_1,\ldots,Z_k}$ after feature selection and $N_\text{res}$ is the resolution parameter of MBTR, i.e., the number of elements used by MBTR to describe each atomic interaction $\text{F}_{k}^{Z_1,\ldots,Z_k}$. 
The normalization over datasets is necessary since not all atomic interactions are present in all datasets. 
Thus, $\iota$ defines the importance of an atomic interaction normalized over datasets which can then be further used for knowledge discover.

\section{Results}\label{sec:results}

In this section, we present and summarize the results of the experiments.

\FloatBarrier
\subsection{Selection of ranking algorithm and cutoff value}\label{ssec:results:rankingandcutoff}

\begin{figure}[ht!]
    \centering
    \includegraphics[width=0.90\textwidth]{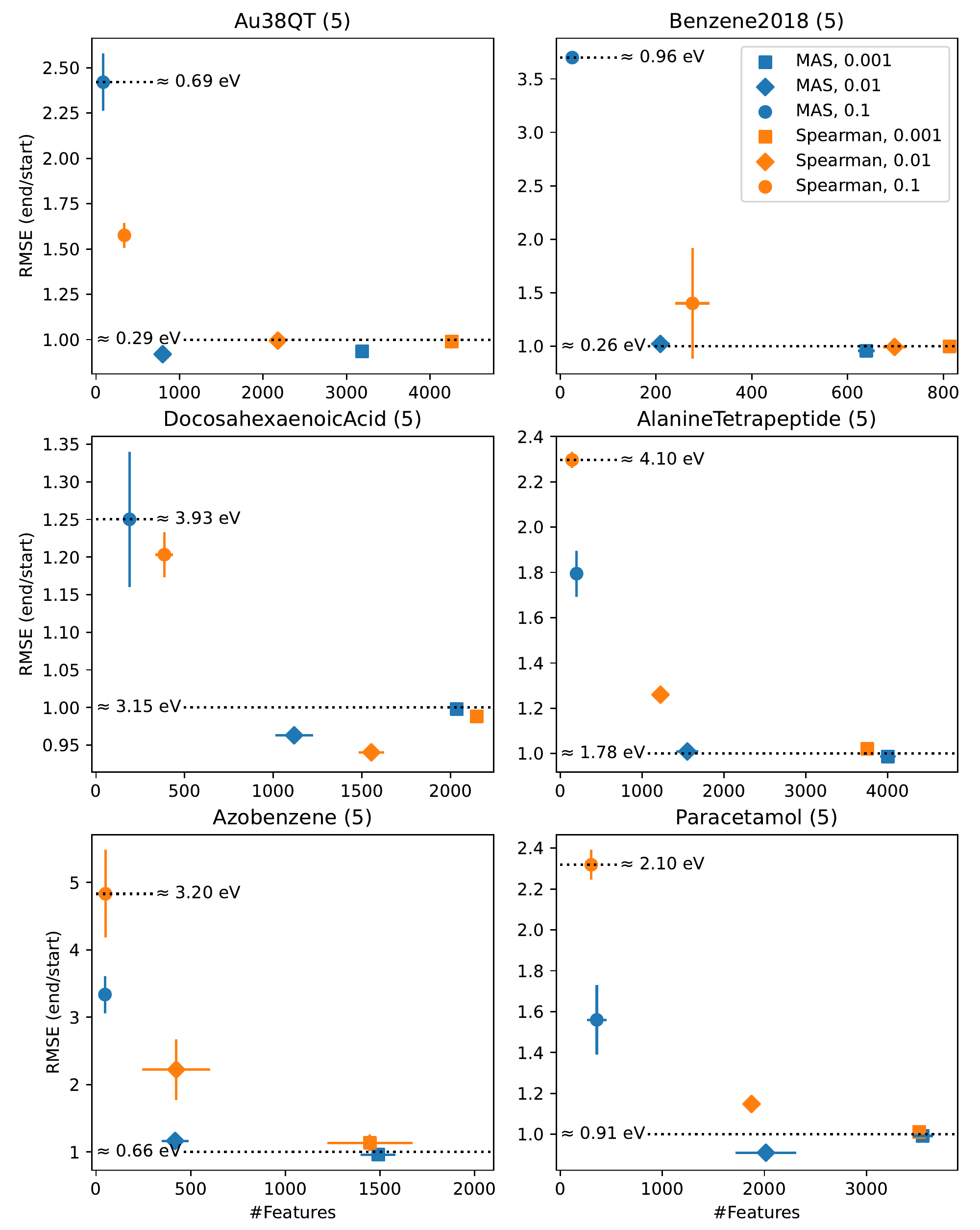}
    \caption{The number of features remaining after feature selection vs. feature selection success measured as model validation RMSE.}
    \label{fig:FEATvsRMSE_1}
\end{figure}

\begin{figure}[ht!]
    \centering
    \includegraphics[width=0.90\textwidth]{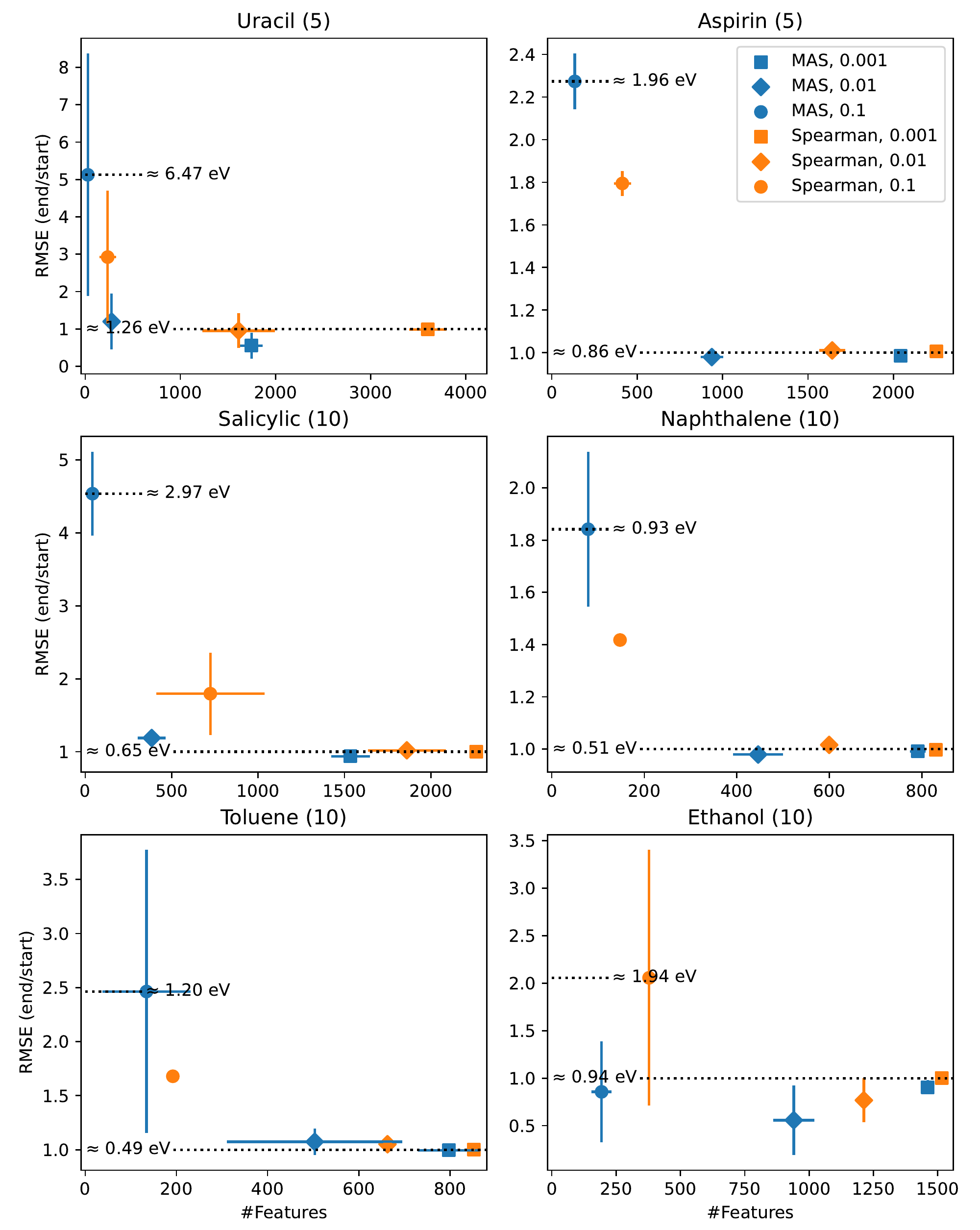}
    \caption{The number of features remaining after feature selection vs. feature selection success measured as model validation RMSE.}
    \label{fig:FEATvsRMSE_2}
\end{figure}

We report the generalizability and FS success with figures where the remaining number of features is on the x-axis and on the y-axis is the validation RMSE after FS divided by the validation RMSE before FS. 
Figure~\ref{fig:FEATvsRMSE_1} and Figure~\ref{fig:FEATvsRMSE_2} show the results for the validation RMSE ratio (error after FS divided by error with a full set of features) as a function of the number of features remaining. 
This error ratio reflects an improvement in regarding model accuracy when it is less than 1. 
Higher values than 1 reflect that the reduced model accuracy is degenerated due to selected subset of features. 

From these figures, we can observe that the cutoff value $0.1$ does not include enough features with either feature scoring algorithm. However, cutoff values $0.01$ and $0.001$ provide at least as low validation errors as the base method but with a significantly reduced number of features. 
The second observation is that \emph{MAS} has a more consistent behavior than \emph{Spearman R} when cutoffs $0.01$ and $0.001$ are considered. 

We used \emph{MAS} with the cutoff $0.01$ to produce figures for all datasets (Figures~\ref{fig:meanmbtr_Au38QT}--\ref{fig:meanmbtr_Ethanol}) where the MBTR (k2+k3) has been laid out with the kept features. 
The k1 part of these figures was removed since, in each case, they were constants that were removed in a preprocessing step. 
The MBTR figures with the selected features for the datasets (Figures~\ref{fig:meanmbtr_Au38QT}--\ref{fig:meanmbtr_Ethanol}) allow one to see what the machine learning model considers to be the most important interactions. 
At the same time, it also allows one to sanity check the function of the feature selection algorithm, provided one has understanding of the involved chemistry. 

Figure~\ref{fig:meanmbtr_Au38QT} presents the MBTR for the most complex object of the twelve datasets, a hybrid metal nanoparticle (a monolayer protected nanocluster). 
The nanoparticle consists of a cluster core composed of gold atoms, a sulfur interface and then 16 hydrocarbon chains in a protective shell around the cluster core. 
The difference of the importances of the features is immediately visible in Figure~\ref{fig:meanmbtr_Au38QT}. 

Then, the choice was between \emph{MAS} and \emph{Spearman R}. 
Observing Figure~\ref{fig:FEATvsRMSE_1} and Figure~\ref{fig:FEATvsRMSE_2} we can see that \emph{MAS} produces lower validation errors in a more consistent manner. 
Due to these findings, the results reported in Section~\ref{ssec:results:selectedfeatures} and Section~\ref{ssec:results:interactionrelevance} are given with \emph{MAS} using cutoff of $0.01$. 

It should be noted, that the validation errors for datasets \emph{Docosahexaenoic Acid} and \emph{AlanineTetrapeptide} presented in Figure~\ref{fig:FEATvsRMSE_1} have a higher validation error than what the other datasets have. 
Both \emph{Docosahexaenoic Acid} and \emph{AlanineTetrapeptide} are long molecule chains with many degrees of freedom, which results in a set of complicated movement during molecular dynamics simulation. 

\FloatBarrier
\subsection{Selected features}\label{ssec:results:selectedfeatures}
\newcommand{\meanmbtrwidth}[0]{\textwidth}
\newcommand{\meanmbtrcaptiontext}[1]{The mean MBTR (k2+k3) for dataset \emph{{#1}} with features selected by Algorithm~\ref{alg:FID} using \emph{MAS} and cutoff of $0.01$}

\begin{figure}[ht!]
    \centering
    \includegraphics[width=\meanmbtrwidth]{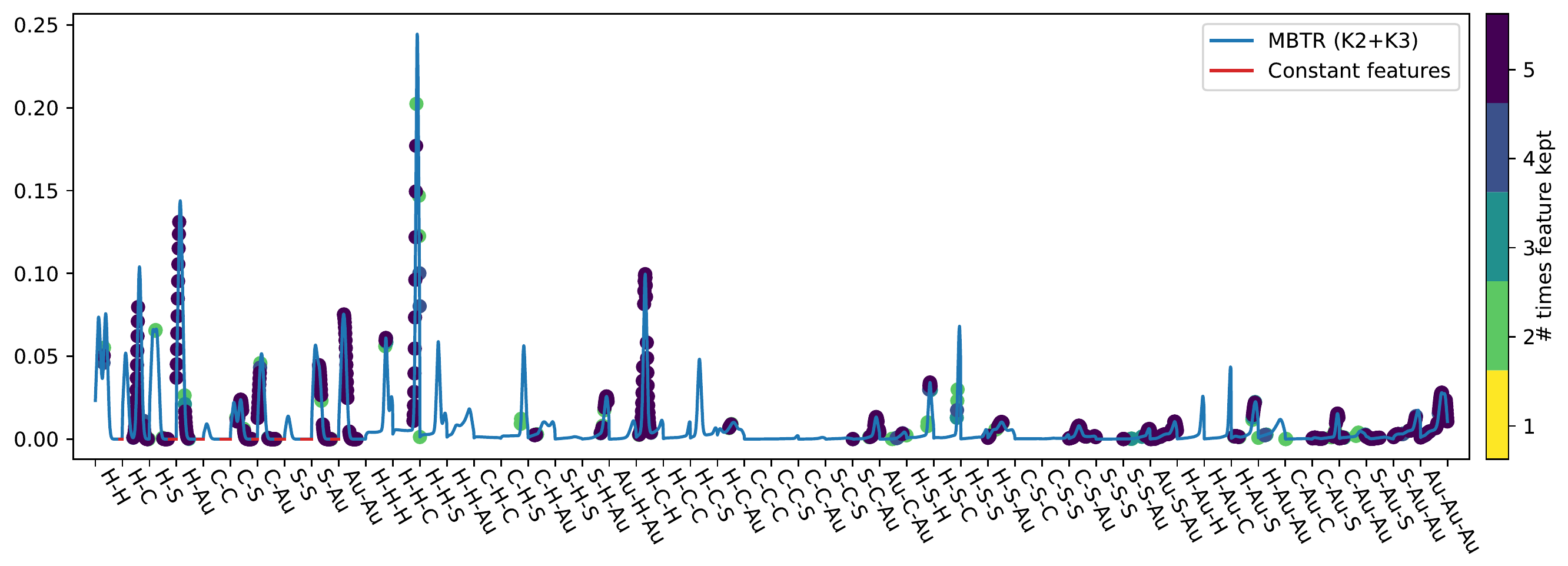}
    \caption{\meanmbtrcaptiontext{Au38QT}}
    \label{fig:meanmbtr_Au38QT}
\end{figure}

\begin{figure}[ht!]
    \centering
    \includegraphics[width=\meanmbtrwidth]{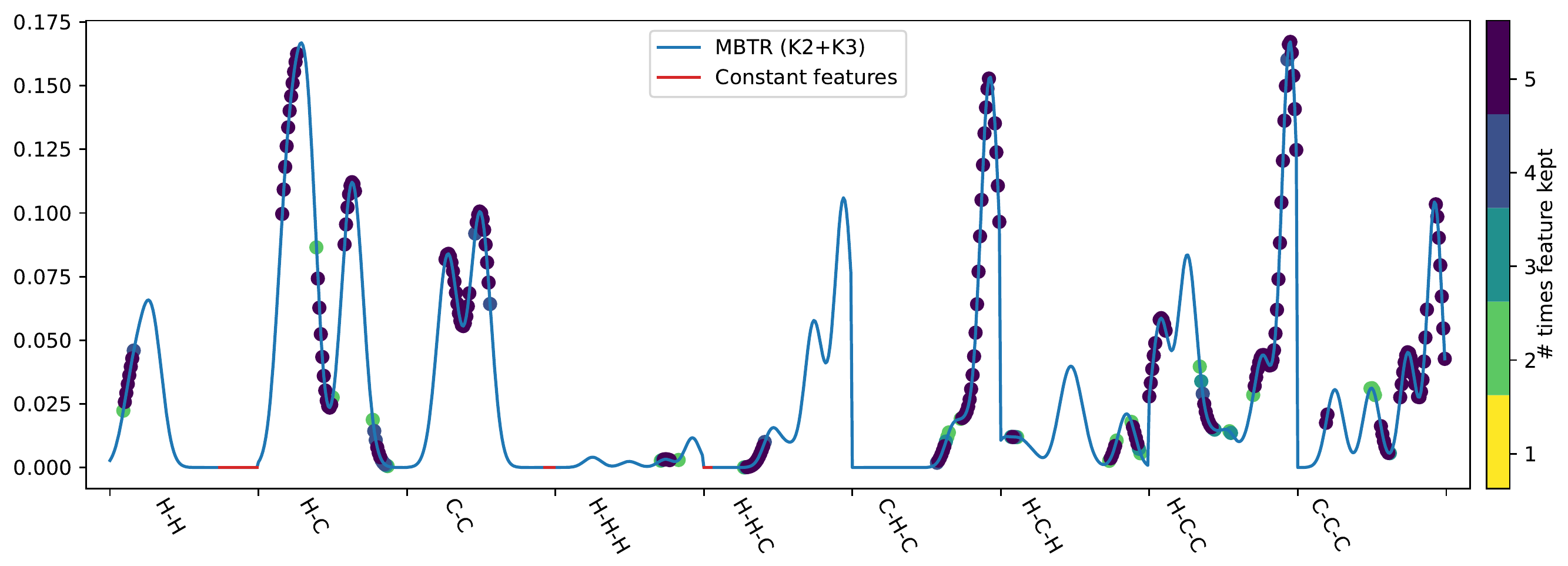}
    \caption{\meanmbtrcaptiontext{Benzene2018}}
    \label{fig:meanmbtr_Benzene2018}
\end{figure}

\begin{figure}[ht!]
    \centering
    \includegraphics[width=\meanmbtrwidth]{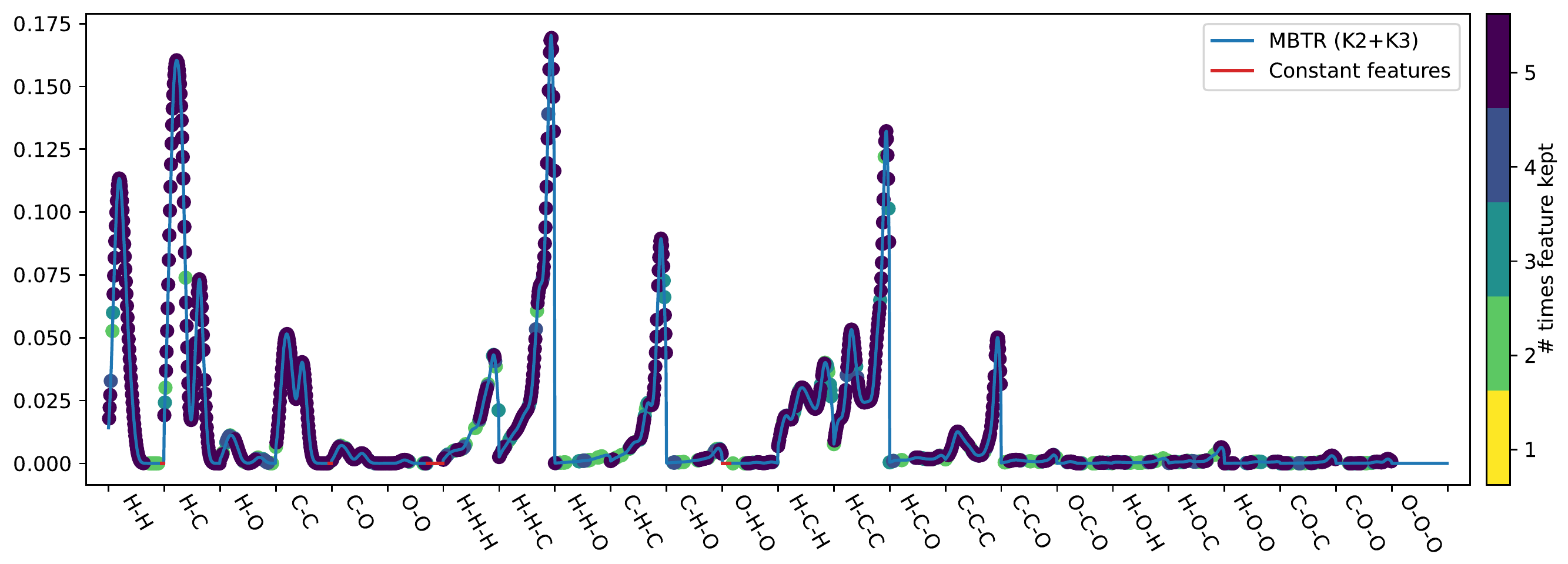}
    \caption{\meanmbtrcaptiontext{DocosahexaenoicAcid}}
    \label{fig:meanmbtr_DocosahexaenoicAcid}
\end{figure}

\begin{figure}[ht!]
    \centering
    \includegraphics[width=\meanmbtrwidth]{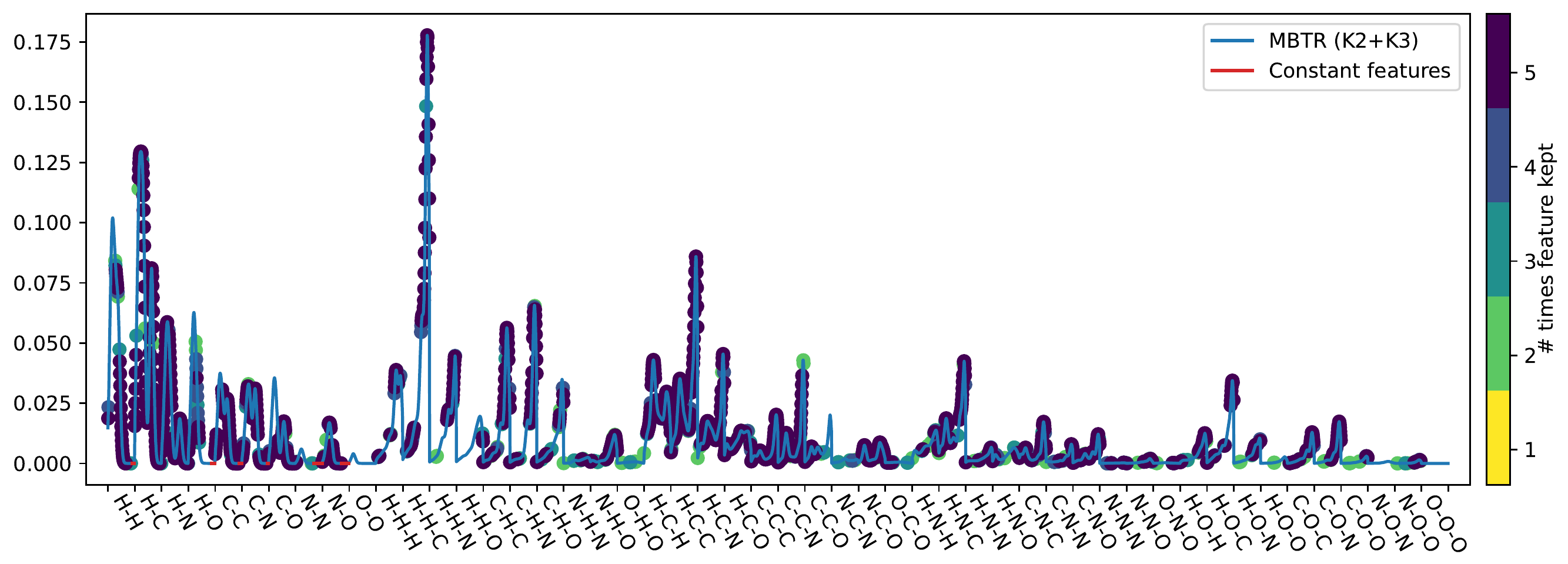}
    \caption{\meanmbtrcaptiontext{AlanineTetrapeptide}}
    \label{fig:meanmbtr_AlanineTetrapeptide}
\end{figure}

\begin{figure}[ht!]
    \centering
    \includegraphics[width=\meanmbtrwidth]{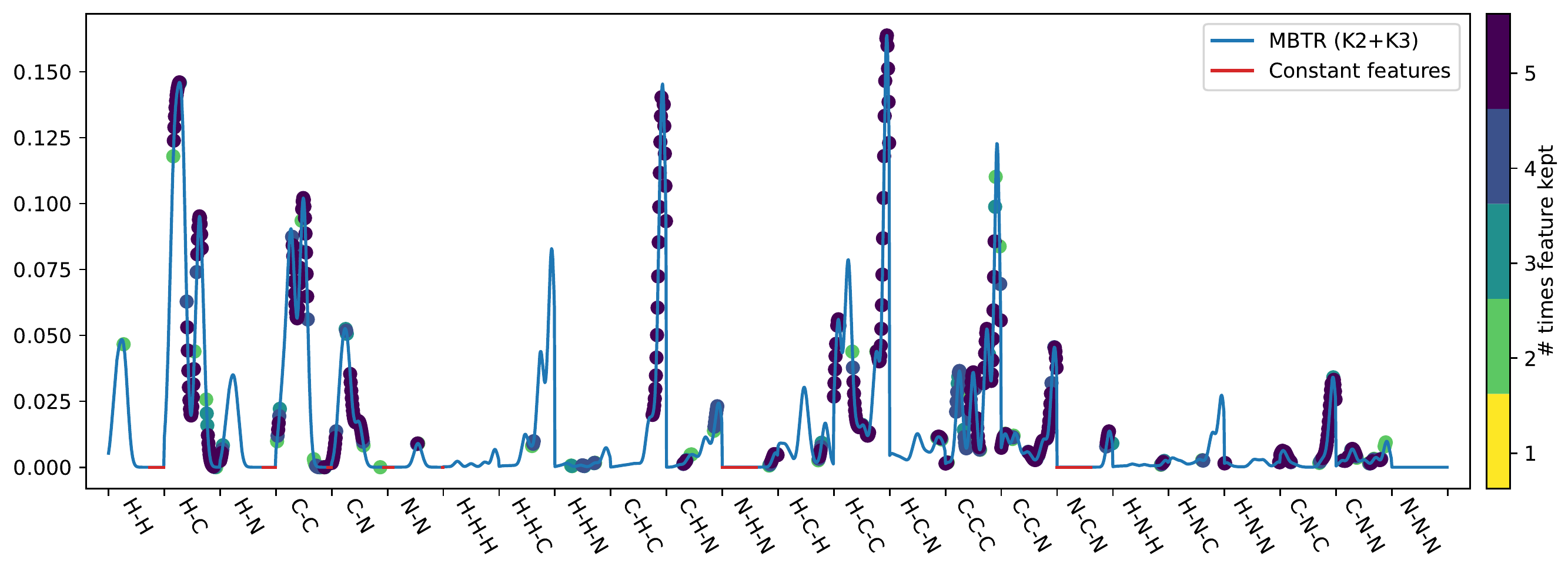}
    \caption{\meanmbtrcaptiontext{Azobenzene}}
    \label{fig:meanmbtr_Azobenzene}
\end{figure}

\begin{figure}[ht!]
    \centering
    \includegraphics[width=\meanmbtrwidth]{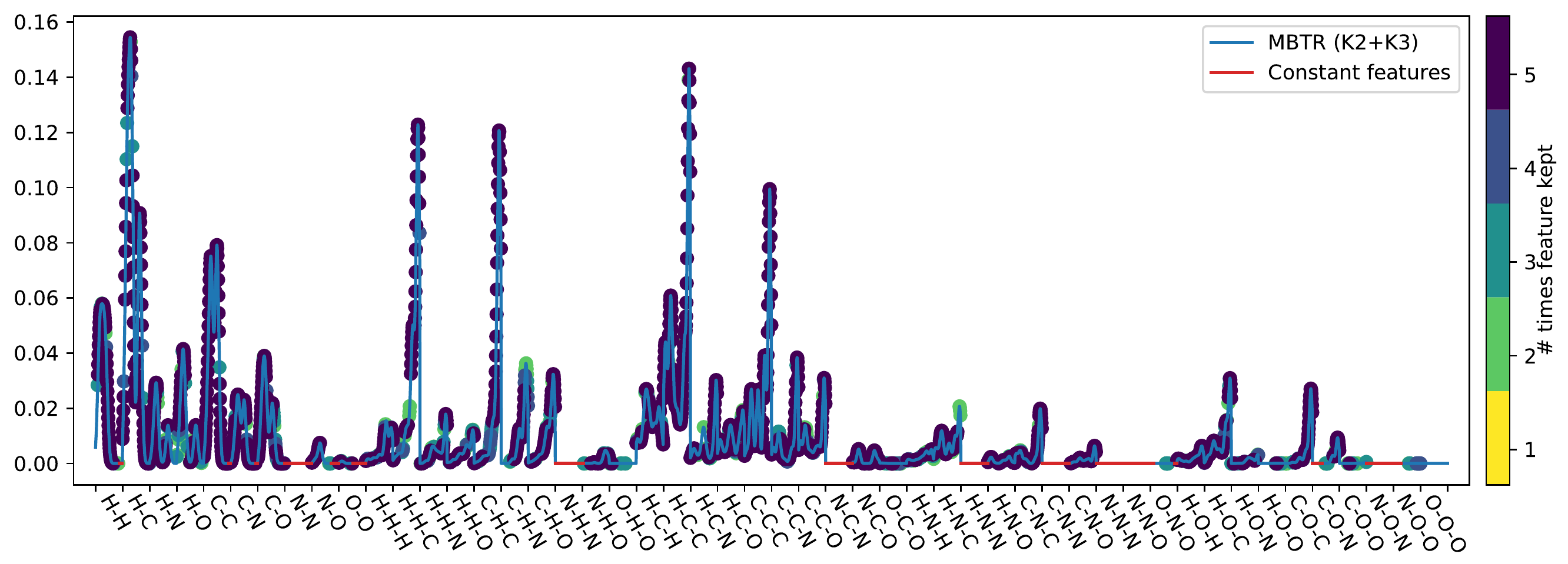}
    \caption{\meanmbtrcaptiontext{Paracetamol}}
    \label{fig:meanmbtr_Paracetamol}
\end{figure}

\begin{figure}[ht!]
    \centering
    \includegraphics[width=\meanmbtrwidth]{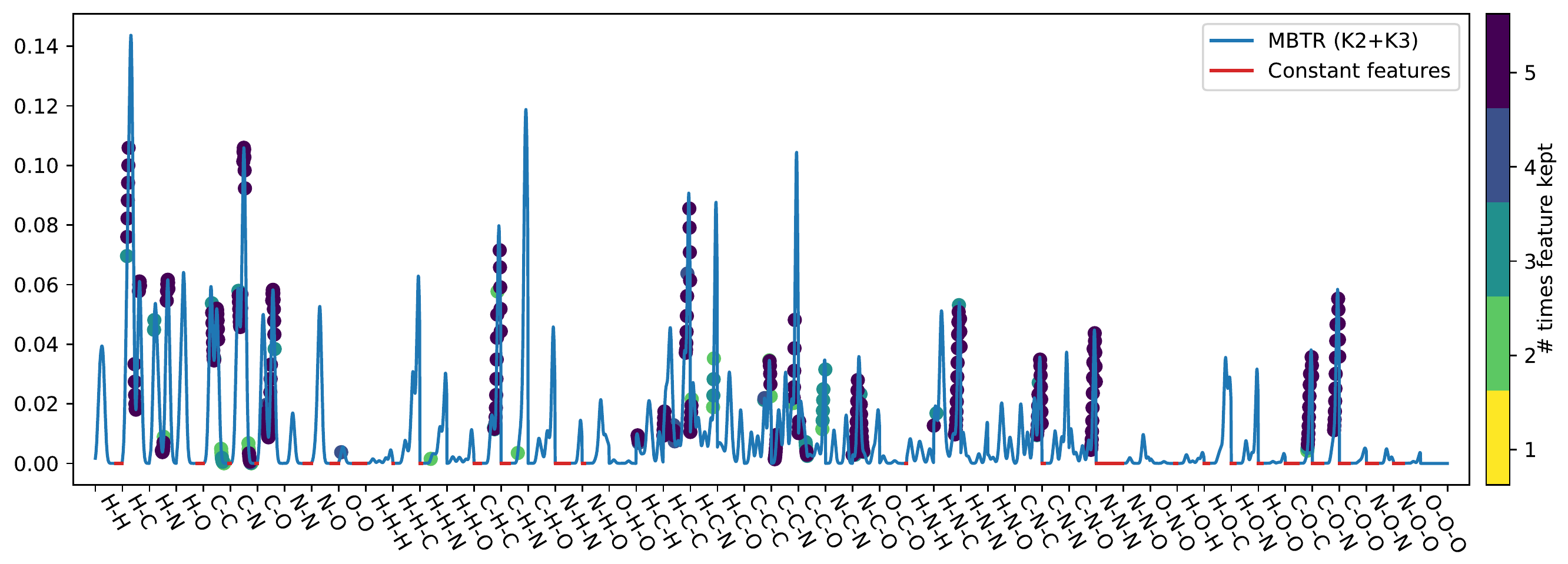}
    \caption{\meanmbtrcaptiontext{Uracil}}
    \label{fig:meanmbtr_Uracil}
\end{figure}

\begin{figure}[ht!]
    \centering
    \includegraphics[width=\meanmbtrwidth]{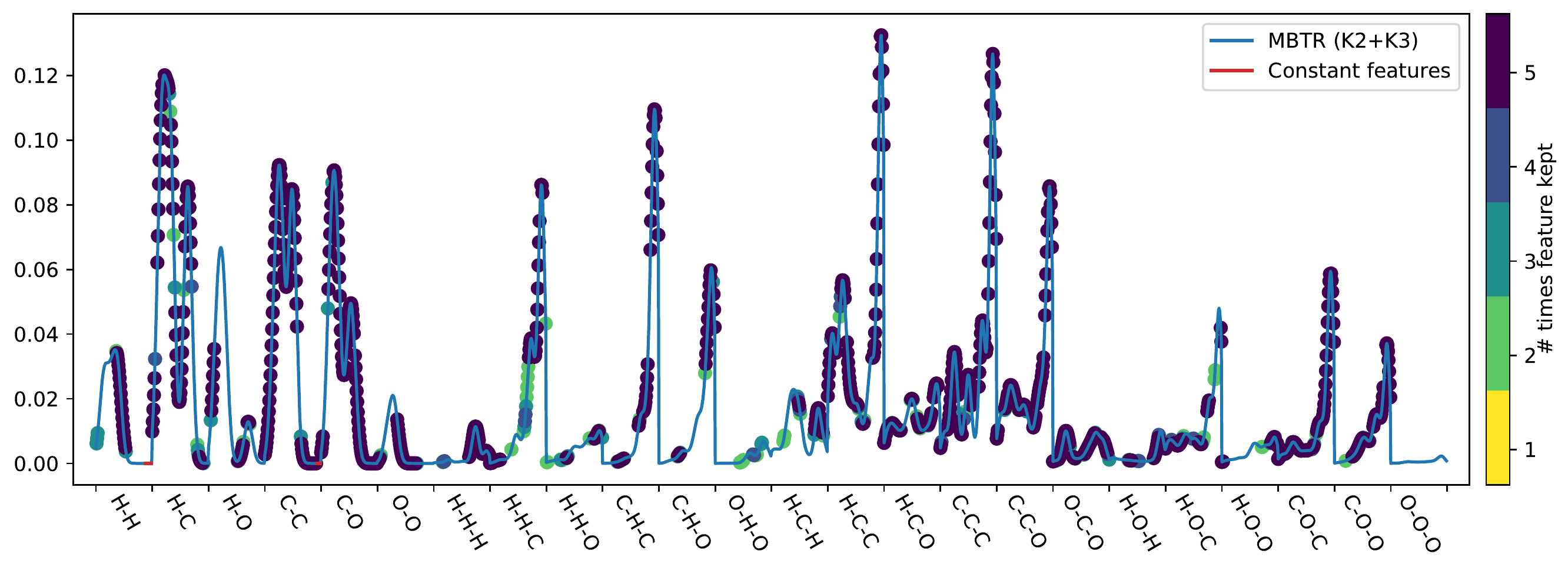}
    \caption{\meanmbtrcaptiontext{Aspirin}}
    \label{fig:meanmbtr_Aspirin}
\end{figure}

\begin{figure}[ht!]
    \centering
    \includegraphics[width=\meanmbtrwidth]{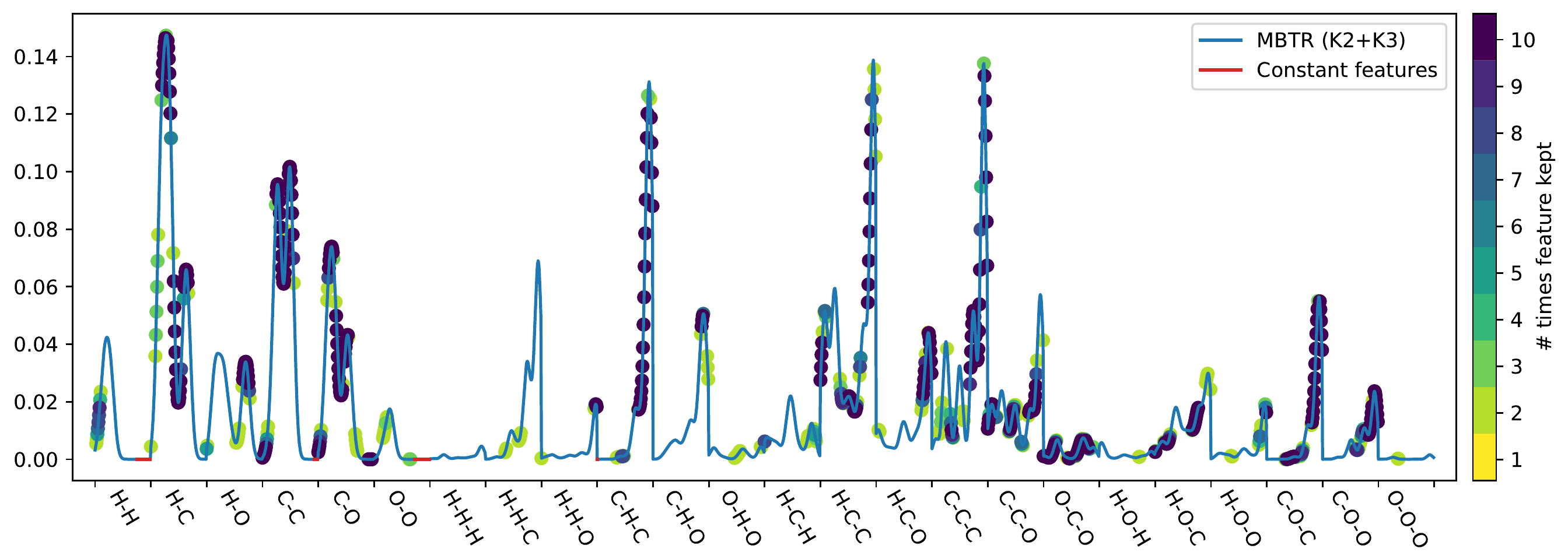}
    \caption{\meanmbtrcaptiontext{Salicylic}}
    \label{fig:meanmbtr_Salicylic}
\end{figure}

\begin{figure}[ht!]
    \centering
    \includegraphics[width=\meanmbtrwidth]{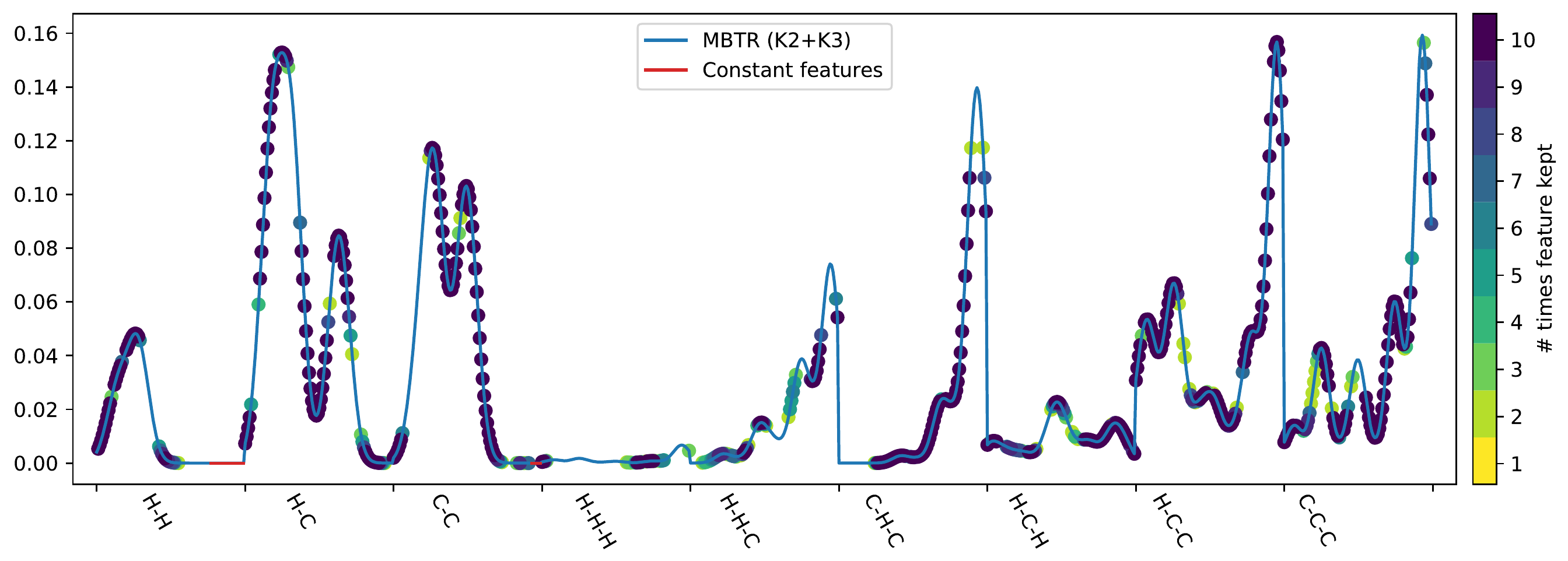}
    \caption{\meanmbtrcaptiontext{Naphthalene}}
    \label{fig:meanmbtr_Naphthalene}
\end{figure}

\begin{figure}[ht!]
    \centering
    \includegraphics[width=\meanmbtrwidth]{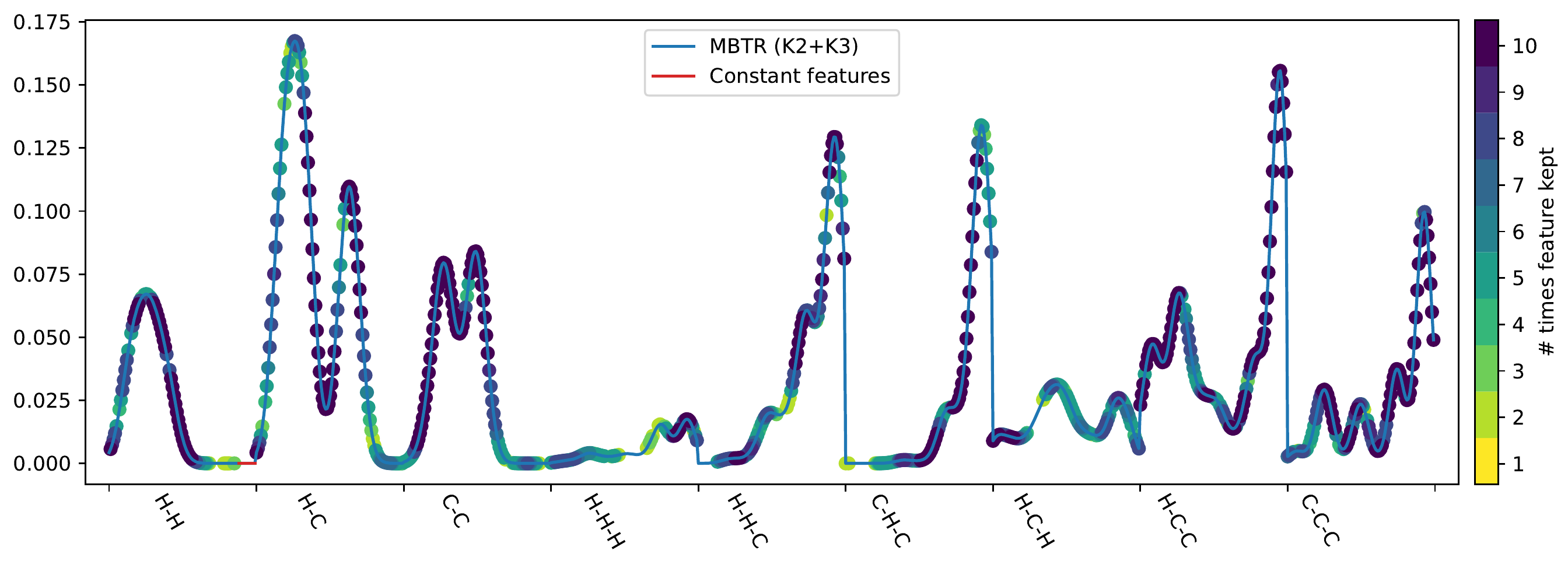}
    \caption{\meanmbtrcaptiontext{Toluene}}
    \label{fig:meanmbtr_Toluene}
\end{figure}

\begin{figure}[ht!]
    \centering
    \includegraphics[width=\meanmbtrwidth]{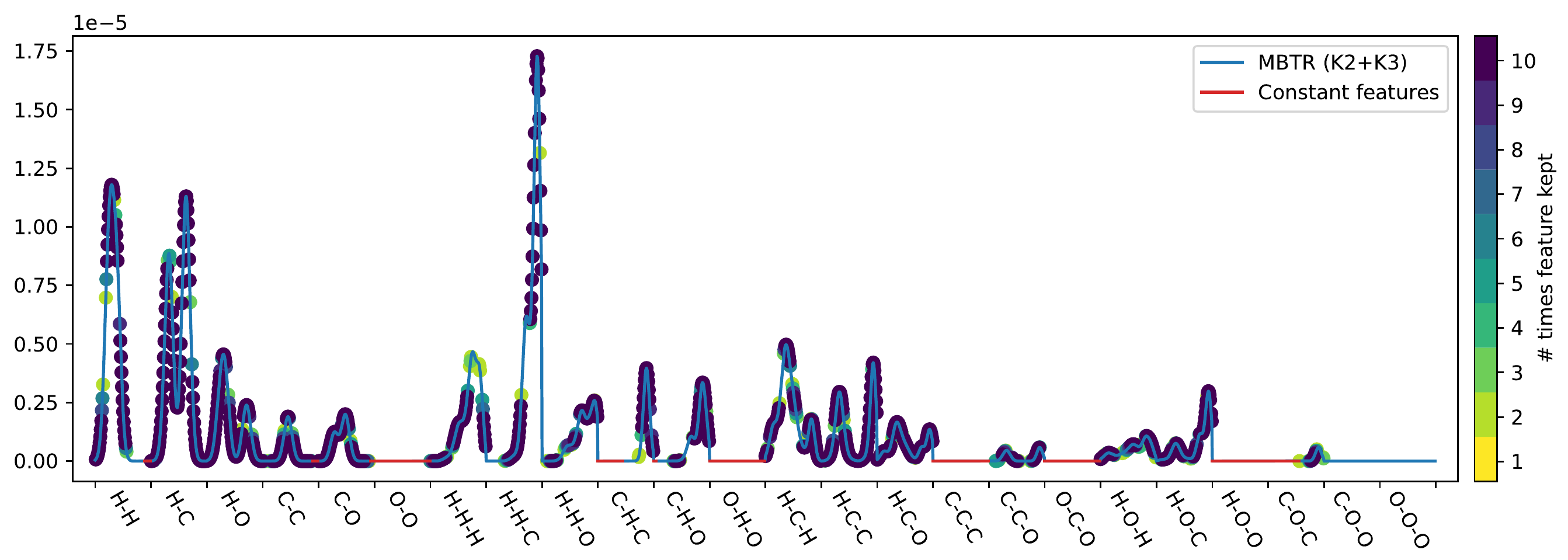}
    \caption{\meanmbtrcaptiontext{Ethanol}}
    \label{fig:meanmbtr_Ethanol}
\end{figure}

Figures~\ref{fig:meanmbtr_Au38QT}--\ref{fig:meanmbtr_Ethanol} contain the mean MBTR for each dataset, along with the selected features (\emph{MAS} and $0.01$) as well as the number of times each selected feature was selected. 
As a general comment, the results presented here are in agreement with chemical intuition. 
The chemical bonds and interactions that one would expect to see are present in the figures. 

It is interesting that the most complex structure, \emph{Au38QT} does not require the highest number of features. 
The most plausible cause of this characteristic is the structurally distinguishable three chemical environments: metallic core, metal--ligand interface and protecting ligands. 
In the simulation data for \emph{Au38QT}, the cluster does break apart at high temperatures but even in those situations, the three main environments of the cluster remain. 
The outcome is that from the point of view of most individual atoms, the local environment remains the same. 
This lessens the need for describing features. 
In addition, the metal--ligand interface of \emph{Au38QT} is one of the defining factors for the stability of the structure \cite{bib:rosalba2019}. 
Therefore, it is expected that the three distinct chemical environments remain visible among the selected features, and as is seen in Figure~\ref{fig:meanmbtr_Au38QT}, they remain visible. 

The rest of the datasets depict organic molecules which are significantly smaller than \emph{Au38QT}. 
This generates a clear distinction between \emph{Au38QT} and the organic molecules. 
In the case of the smaller molecules there are fewer atoms in total, which means that each atom has relatively larger contribution to the potential energy of the molecule. 
In addition to each atom having a relatively larger contribution, the smaller molecules themselves have less moving parts and thus potentially have less movement and higher stability, when compared to \emph{Au38QT}. 

The remaining features and the MBTR descriptors of the organic molecules are presented in Figures~\ref{fig:meanmbtr_Benzene2018}--\ref{fig:meanmbtr_Ethanol}. 
As mentioned earlier in this section, the organic molecules have generally demanded a high number of features. 
In the case of \emph{Benzene} in dataset \emph{Benzene2018}, very few features are needed since the molecule is a stable ring. 
In the cases of \emph{Docosahexaenoic Acid} and \emph{AlanineTetrapeptide} in  Figures~\ref{fig:meanmbtr_DocosahexaenoicAcid} and \ref{fig:meanmbtr_AlanineTetrapeptide}, both are long chains with multiple and varying interactions. 
This means that each individual atom sees numerous local environments and that the molecule chain, as it can twist and turn or remain straight or anything in between, creates a complex potential energy surface. 
The complexity of the potential energy surface then requires a large number of features to work with. 

Among the remaining molecules, \emph{Azobenzene} in Figure~\ref{fig:meanmbtr_Azobenzene}, \emph{Uracil} in Figure~\ref{fig:meanmbtr_Uracil}, \emph{Salicylic Acid} in Figure~\ref{fig:meanmbtr_Salicylic} and \emph{Toluene} in Figure~\ref{fig:meanmbtr_Toluene} are similar to \emph{Benzene} in the way that the molecules have stable structures and not a lot of movement. 
This results in them requiring fewer number of features. 
On the other hand are molecules \emph{Paracetamol} in Figure~\ref{fig:meanmbtr_Paracetamol}, \emph{Aspirin} in Figure~\ref{fig:meanmbtr_Aspirin}, \emph{Naphthalene} in Figure~\ref{fig:meanmbtr_Naphthalene} and \emph{Ethanol} in Figure~\ref{fig:meanmbtr_Ethanol} which required a larger number of features to be kept. 
Each (except \emph{Naphthalene}) has molecular groups which are able to move, which may be the reason for their need for features. 
\emph{Naphthalene} is an unexpected case, as the molecule does not have moving parts. 
The structure of \emph{Naphthalene} is however clearly visible in the selected features, the distances and angles have remained as expected. 

The smaller organic molecules surprised with the number of features they required. 
A potential point of optimization would be to optimize the cutoff parameter to \emph{Au38QT} and the organic molecules separately, instead of optimizing both at the same time. 
The expectation would be that due to the difference in chemistry, the cutoff points would differ in fully optimized situation. 
It would then affect the selected number of features.

\FloatBarrier
\subsection{Interaction relevance}\label{ssec:results:interactionrelevance}

\begin{figure}[b!]
    \centering
    \includegraphics[width=\textwidth]{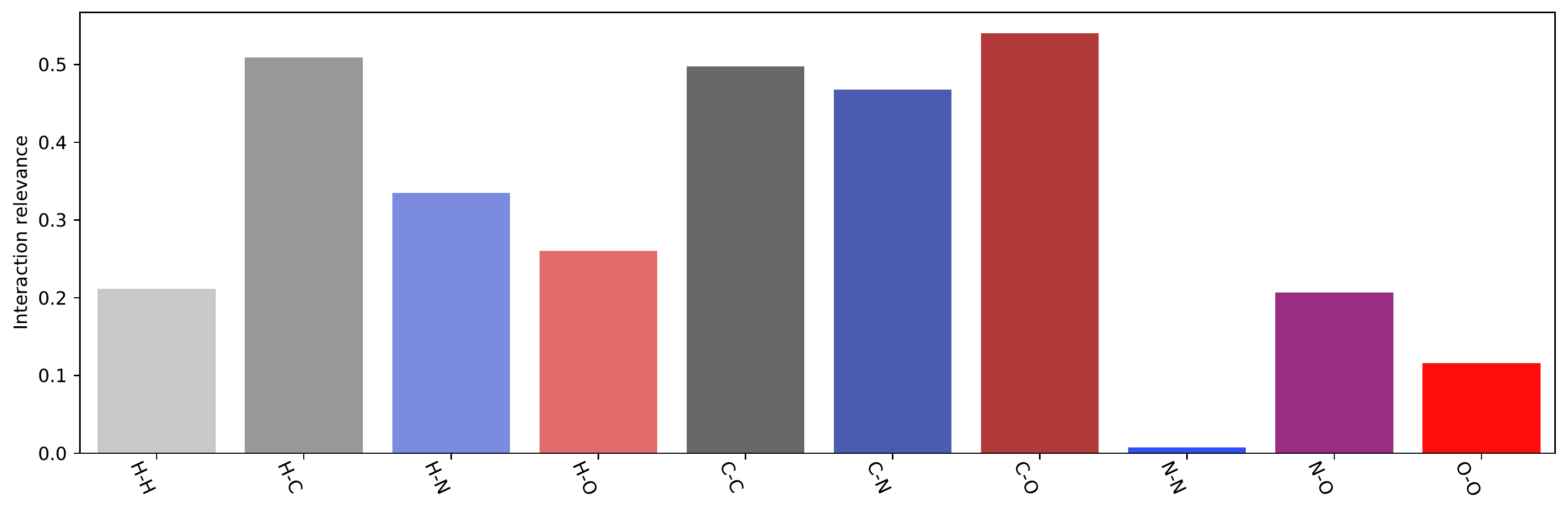}
    \caption{Atomic interaction presence for all organic molecule datasets with MBTR k2 (inverse distances). Interactions with an empty bar were not present in the datasets after feature selection.}
    \label{fig:interactionPrecense_organic_K2}
\end{figure}

\begin{figure}[t!]
    \centering
    \includegraphics[width=\textwidth]{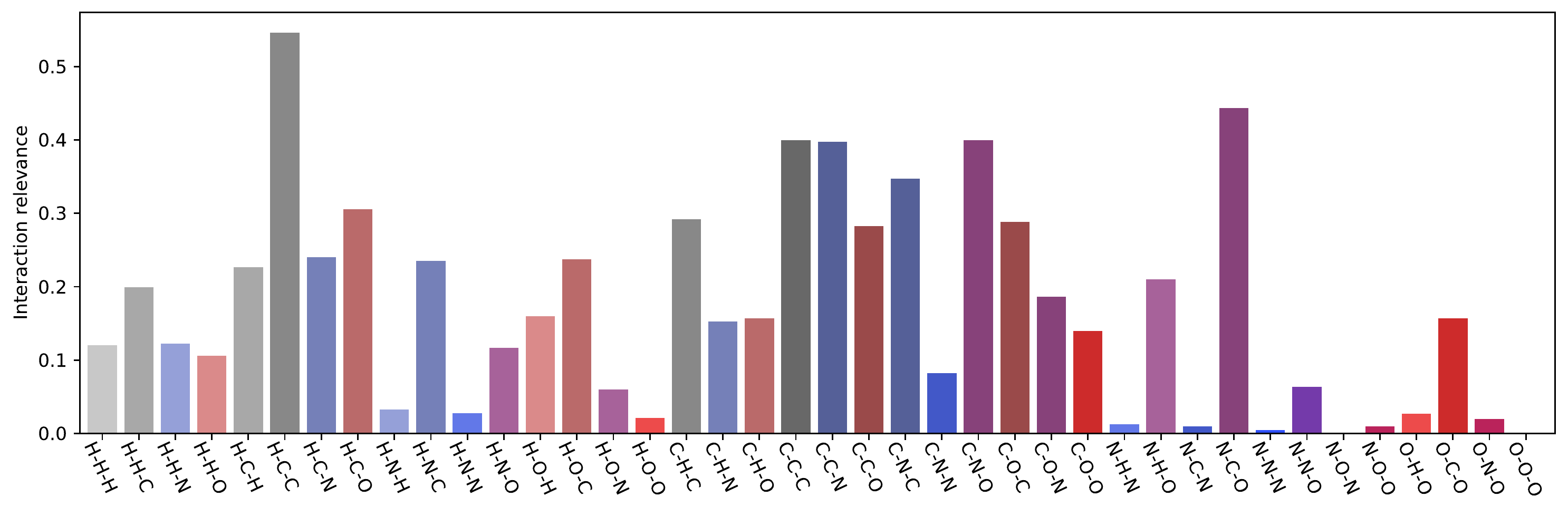}
    \caption{Atomic interaction presence for all organic molecule datasets with MBTR k3 (angle between a triple). Interactions with an empty bar were not present in the datasets after feature selection.}
    \label{fig:interactionPrecense_organic_K3}
\end{figure}

\begin{figure}[t!]
    \centering
    \includegraphics[width=\textwidth]{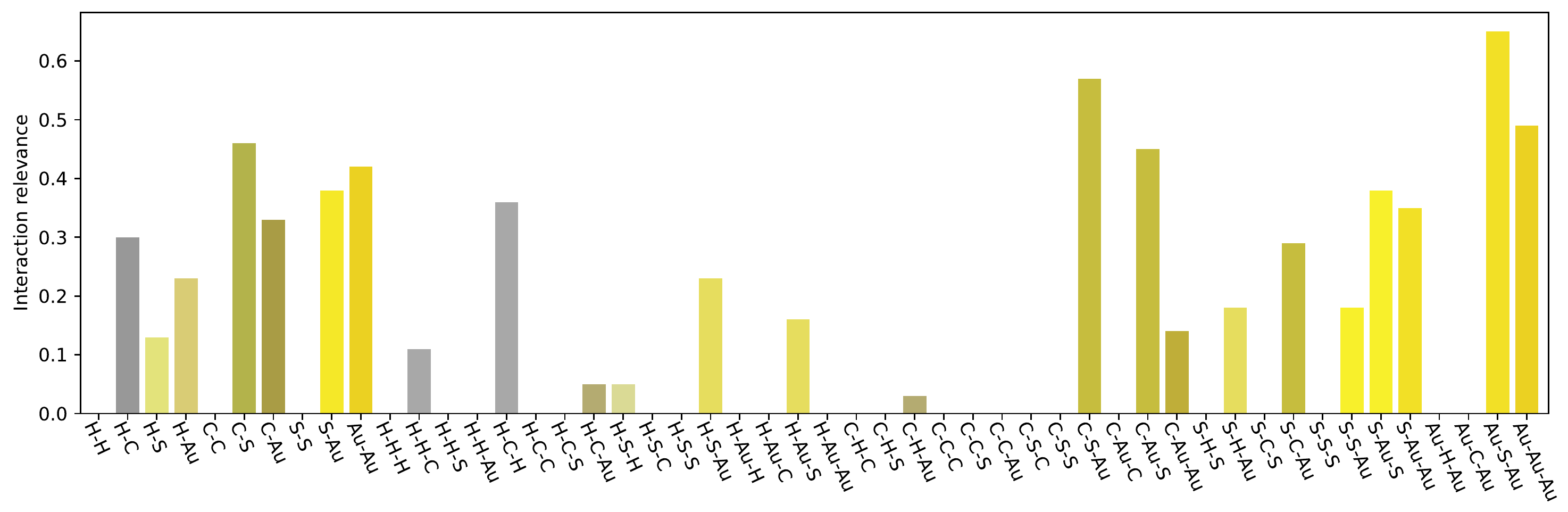}
    \caption{Atomic interaction presence for dataset \emph{Au38QT} with MBTR k2+k3. Interactions with an empty bar were not present in the datasets after feature selection.}
    \label{fig:interactionPrecense_cluster_K23}
\end{figure}

The results here are a summarization of the chemical interactions present in the studied atomic structures. 
The results for the interaction relevance are presented in Figures~\ref{fig:interactionPrecense_organic_K2} and \ref{fig:interactionPrecense_organic_K3} for organic molecules and Figure~\ref{fig:interactionPrecense_cluster_K23} for \emph{Au38QT}. 
The results were split into figures this way due to the difference between the chemistry of an organic molecule and the chemistry of a hybrid nanoparticle. 

Considering the importance of carbon in organic molecules, it is not a surprise that interactions with carbon were the most important atomic bonds in the datasets in Figure~\ref{fig:interactionPrecense_organic_K2}. 
The apparent lack of importance for N-N bond distance is explained by the N-N bond in \emph{Azobenzene} being a double bond and said N-N bond being the only example of N-N bond present in the datasets. 
It is an especially stable bond, thus in the eyes of feature selection, N-N bonds had little importance to the potential energy. 
We would expect the N-N bar to have more presence had the datasets contained N-N bonds that were more flexible. 

From the angles between atoms presented in Figure~\ref{fig:interactionPrecense_organic_K3}, the angle H-C-C has the highest relevance. 
This is again expected based on the nature of organic molecules. 
The same trend as in Figure~\ref{fig:interactionPrecense_organic_K2} is seen in Figure~\ref{fig:interactionPrecense_organic_K3}, interactions containing carbon are ranked the most relevant. 

The results for the \emph{Au38QT} are presented in Figure~\ref{fig:interactionPrecense_cluster_K23}. 
One can clearly see the dominant effects of gold and sulfur in the interaction relevances. 
This is again expected based on knowledge of chemistry. 
An interesting note is that the highest relevances are given to gold--sulfur--gold and carbon--sulfur--gold. 
In other words, the interface between the gold core and the ligands. 
Au-S-Au and Au-S-C angles are generally around $90^\circ$, hence they have a clear local energy minimum configuration. Deviation from this angle is expected to have visible effect on the potential energy, which explains why they get high relevance. 

While the results presented here are mostly a confirmation that the proposed feature selection method works as intended, results like these could be used in the priorization of atom-atom parametrization. 
In other word, it is an automated way to print out what the machine learning model in combination with the feature selection method has seen as the most relevant area of interest.

\FloatBarrier
%==========================================================================
%==========================================================================
\section{Conclusions}\label{sec:conclusions}

The proposed feature selection algorithm was developed and tested with datasets based on molecular dynamics simulations on organic molecules and a hybrid nanoparticle. 
The Feature Importance Detector requires only a single hyperparameter when used with full EMLM (all of input data selected as reference points) and it was validated by using it with simulation data and the analysis of the selected features. 
The hyperparameter of FID was first searched by comparing Spearman R and MAS to each other. 
Other potential feature scoring algorithms were not used here due to the testing made earlier by Linja et al. in \cite{bib:linja2023}. 
The FID was then used to select features from the used datasets. 
The results were analyzed with the use of domain knowledge. 
Finally, interaction relevance score was used to present another way to infer information from the combination of feature selection method, machine learning model, descriptor and molecular dynamics simulation data. 

The feature relevance metric was used to summarize the chemical information, as given by the molecular dynamics data, descriptor, EMLM and the proposed feature selection algorithm. 
It was used to aggregate the information in the organic molecules and single out the \emph{Au38QT}. 
We conclude that the proposed feature selection algorithm functions as intended, as the results it gave were analyzable and verifiable through domain knowledge. 

As this work utilized a single descriptor, one future work would be to compare the result analysis to the result analysis of other descriptors. 
Jäger et al. \cite{bib:jager2018} showed that there are situations where local descriptors are better than global descriptors. 
This naturally leads to a future work where local and global descriptors are used simultaneously with FID to see the relative importances of both. 
An additional avenue of research would be to focus more on different nanoclusters and attempt to analyze whether the features or the interaction relevances have common elements among the various nanoclusters. 
Another option would be to work on gaining a model from descriptor to 3D structure. 

%==========================================================================
%==========================================================================
%==========================================================================
\medskip

\textbf{Acknowledgements} \par 
This work has been supported by the Academy of Finland through the project 351579 (MLNovCat). 
We acknowledge grants of computer capacity from the Finnish Grid and Cloud Infrastructure (\textbf{FCCI}; persistent identifier urn:nbn:fi:research-infras-2016072533). 

\section*{Conflict of interest}

The authors declare no conflict of interest.

\medskip

\bibliographystyle{unsrt}
\bibliography{references.bib}

\newpage
\section*{Appendix A}
\pagenumbering{roman}
\setcounter{page}{1}
\newcommand{\MolFigSubFigW}{0.395\textwidth}

\begin{figure}[ht!]
\centering
\begin{subfigure}{\MolFigSubFigW}
    \includegraphics[width=\textwidth]{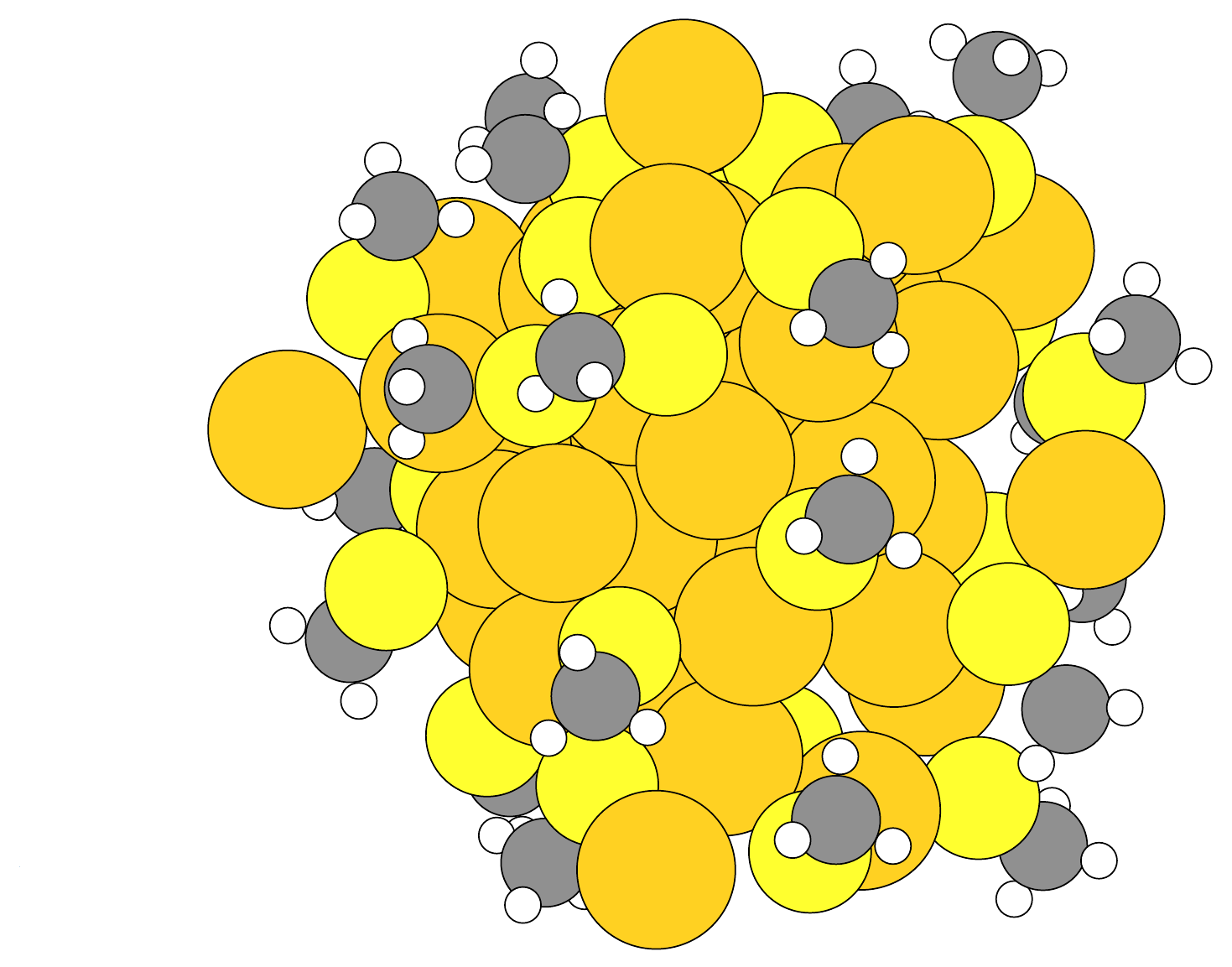}
    \caption{\emph{Au38QT}}
    \label{fig:MolecularFigures:Au38QT}
\end{subfigure}
\begin{subfigure}{\MolFigSubFigW}
    \includegraphics[width=\textwidth]{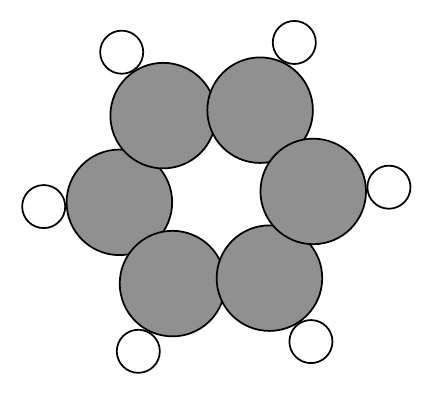}
    \caption{\emph{Benzene}}
    \label{fig:MolecularFigures:Benzene}
\end{subfigure}
\\
\begin{subfigure}{\MolFigSubFigW}
    \includegraphics[width=\textwidth]{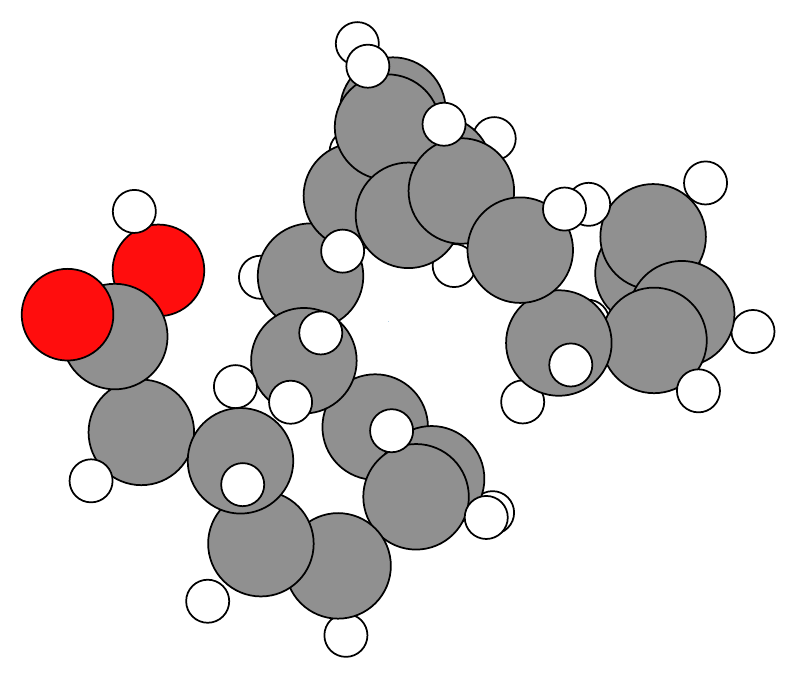}
    \caption{\emph{Docosahexaenoic Acid}}
    \label{fig:MolecularFigures:DocosahexaenoicAcid}
\end{subfigure}
\begin{subfigure}{\MolFigSubFigW}
    \includegraphics[width=\textwidth]{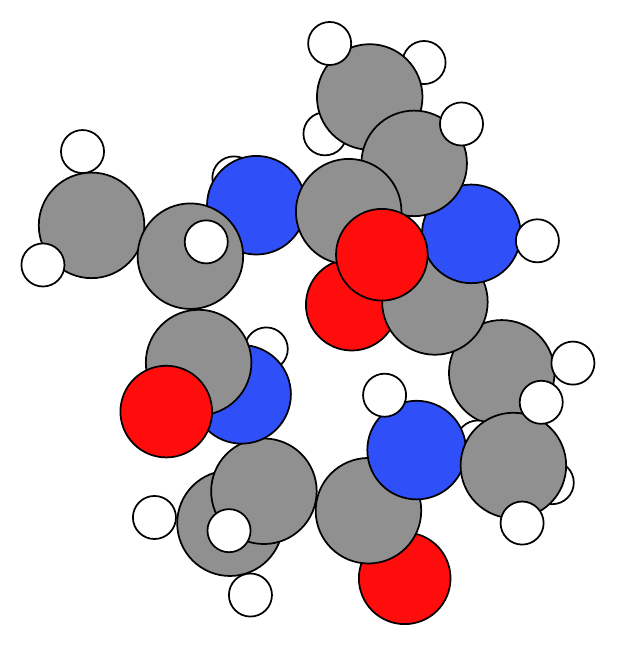}
    \caption{\emph{Alanine Tetrapeptide}}
    \label{fig:MolecularFigures:AlanineTetrapeptide}
\end{subfigure}
\\
\begin{subfigure}{\MolFigSubFigW}
    \includegraphics[width=\textwidth]{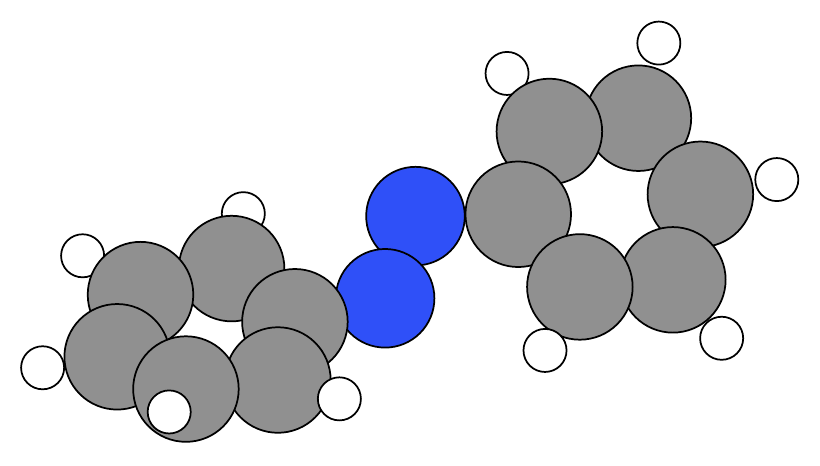}
    \caption{\emph{Azobenzene}}
    \label{fig:MolecularFigures:Azobenzene}
\end{subfigure}
\begin{subfigure}{\MolFigSubFigW}
    \includegraphics[width=\textwidth]{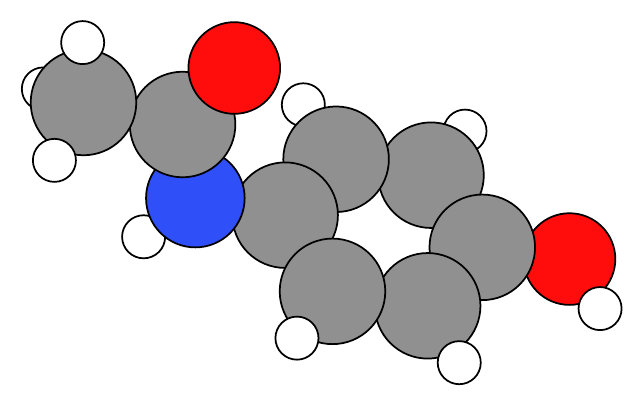}
    \caption{\emph{Paracetamol}}
    \label{fig:MolecularFigures:Paracetamol}
\end{subfigure}
\caption{Example figures of the molecular structure of the first six datasets}
\label{fig:MolecularFigures:1}
\end{figure}

\begin{figure}[ht!]
\centering
\begin{subfigure}{\MolFigSubFigW}
    \includegraphics[width=\textwidth]{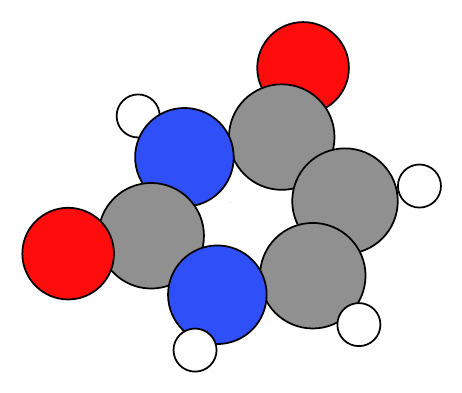}
    \caption{\emph{Uracil}}
    \label{fig:MolecularFigures:Uracil}
\end{subfigure}
\begin{subfigure}{\MolFigSubFigW}
    \includegraphics[width=\textwidth]{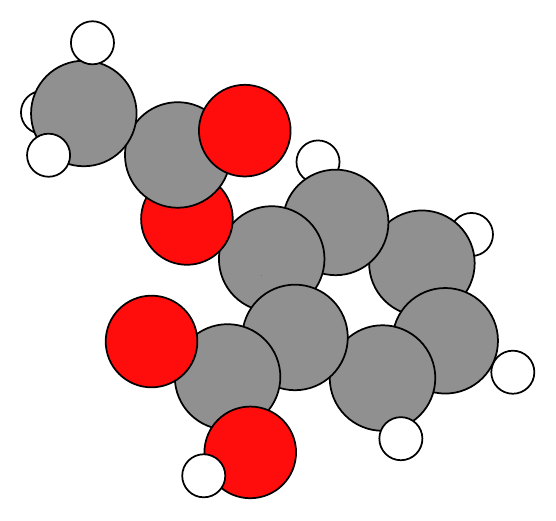}
    \caption{\emph{Aspirin}}
    \label{fig:MolecularFigures:Aspirin}
\end{subfigure}
\\
\begin{subfigure}{\MolFigSubFigW}
    \includegraphics[width=\textwidth]{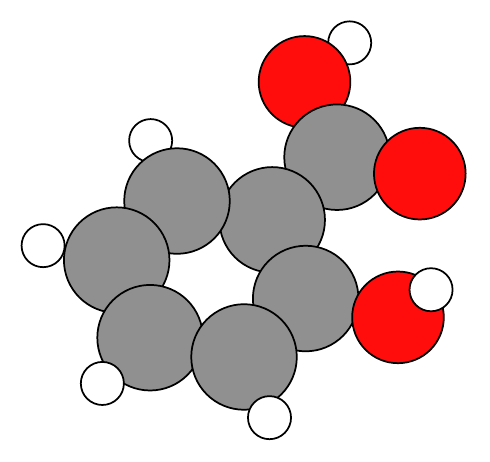}
    \caption{\emph{Salicylic Acid}}
    \label{fig:MolecularFigures:SalicylicAcid}
\end{subfigure}
\begin{subfigure}{\MolFigSubFigW}
    \includegraphics[width=\textwidth]{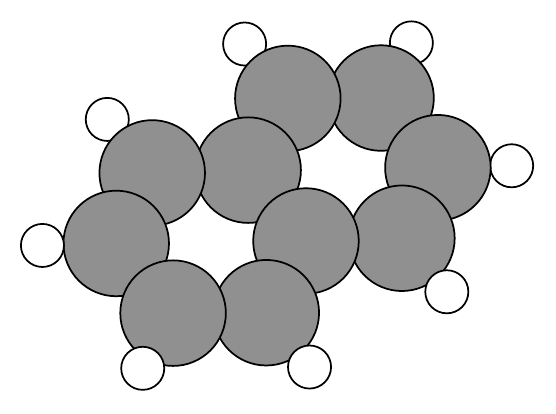}
    \caption{\emph{Naphthalene}}
    \label{fig:MolecularFigures:Naphthalene}
\end{subfigure}
\\
\begin{subfigure}{\MolFigSubFigW}
    \includegraphics[width=\textwidth]{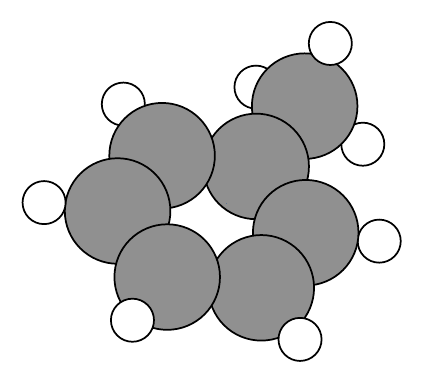}
    \caption{\emph{Toluene}}
    \label{fig:MolecularFigures:Toluene}
\end{subfigure}
\begin{subfigure}{\MolFigSubFigW}
    \includegraphics[width=\textwidth]{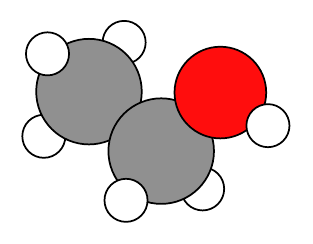}
    \caption{\emph{Ethanol}}
    \label{fig:MolecularFigures:Ethanol}
\end{subfigure}
\caption{Example figures of the molecular structure of the second six datasets}
\label{fig:MolecularFigures:2}
\end{figure}

\end{document}